\documentclass[preprint2]{aastex62}
\def\Mgii{Mg\,{\sc ii}}
\def\Civ{C\,{\sc iv}}
\def\Siv{S\,{\sc iv}}
\def\Nv{N\,{\sc v}}

\def\Cii{C\,{\sc ii}}

\shorttitle{Exploring Reionization-Era Quasars}
\shortauthors{Wang et al.}

\begin{document}
\title{Exploring Reionization-Era Quasars III: Discovery of 16 Quasars at $6.4\lesssim z \lesssim 6.9$ with DESI Legacy Imaging Surveys and UKIRT Hemisphere Survey and Quasar Luminosity Function at $z\sim6.7$}

\correspondingauthor{Feige Wang}
\email{fgwang@physics.ucsb.edu}

\author[0000-0002-7633-431X]{Feige Wang}
\affil{Department of Physics, University of California, Santa Barbara, CA 93106-9530, USA}
\affil{Kavli Institute for Astronomy and Astrophysics, Peking University, Beijing 100871, China}
\affil{Steward Observatory, University of Arizona, 933 North Cherry Avenue, Tucson, AZ 85721, USA}

\author[0000-0001-5287-4242]{Jinyi Yang}
\affil{Steward Observatory, University of Arizona, 933 North Cherry Avenue, Tucson, AZ 85721, USA}
\affil{Kavli Institute for Astronomy and Astrophysics, Peking University, Beijing 100871, China}

\author[0000-0003-3310-0131]{Xiaohui Fan}
\affil{Steward Observatory, University of Arizona, 933 North Cherry Avenue, Tucson, AZ 85721, USA}

\author[0000-0002-7350-6913]{Xue-Bing Wu}
\affil{Kavli Institute for Astronomy and Astrophysics, Peking University, Beijing 100871, China}
\affil{Department of Astronomy, School of Physics, Peking University, Beijing 100871, China}

\author[0000-0002-5367-8021]{Minghao Yue}
\affil{Steward Observatory, University of Arizona, 933 North Cherry Avenue, Tucson, AZ 85721, USA}

\author[0000-0001-6239-3821]{Jiang-Tao Li}
\affil{Department of Astronomy, University of Michigan, 311 West Hall, 1085 S. University Ave, Ann Arbor, MI, 48109-1107, USA}

\author[0000-0002-1620-0897]{Fuyan Bian}
\affil{European Southern Observatory, Alonso de C\'ordova 3107, Casilla 19001, Vitacura, Santiago 19, Chile}

\author[0000-0003-4176-6486]{Linhua Jiang}
\affil{Kavli Institute for Astronomy and Astrophysics, Peking University, Beijing 100871, China}

\author[0000-0002-2931-7824]{Eduardo Ba\~nados}
\affil{The Observatories of the Carnegie Institution for Science, 813 Santa Barbara Street, Pasadena, California 91101, USA}

\author[0000-0002-4544-8242]{Jan-Torge Schindler}
\affil{Steward Observatory, University of Arizona, 933 North Cherry Avenue, Tucson, AZ 85721, USA}

\author[0000-0002-3632-2015]{Joseph R. Findlay}
\affil{University of Wyoming, Physics \& Astronomy 1000 E. University, Dept 3905 Laramie, WY 82071, USA}

\author[0000-0003-0821-3644]{Frederick B. Davies}
\affil{Department of Physics, University of California, Santa Barbara, CA 93106-9530, USA}

\author[0000-0002-2662-8803]{Roberto Decarli}
\affil{INAF--Osservatorio di Astrofisica e Scienza dello Spazio, via Gobetti 93/3, I-40129, Bologna, Italy}

\author[0000-0002-6822-2254]{Emanuele P. Farina}
\affil{Department of Physics, University of California, Santa Barbara, CA 93106-9530, USA}

\author{Richard Green}
\affil{Steward Observatory, University of Arizona, 933 North Cherry Avenue, Tucson, AZ 85721, USA}

\author[0000-0002-7054-4332]{Joseph F. Hennawi}
\affil{Department of Physics, University of California, Santa Barbara, CA 93106-9530, USA}
\affil{Max Planck Institut f\"ur Astronomie, K\"onigstuhl 17, D-69117, Heidelberg, Germany}

\author[0000-0003-4955-5632]{Yun-Hsin Huang}
\affil{Steward Observatory, University of Arizona, 933 North Cherry Avenue, Tucson, AZ 85721, USA}

\author[0000-0002-5941-5214]{Chiara Mazzuccheli}
\affil{Max Planck Institut f\"ur Astronomie, K\"onigstuhl 17, D-69117, Heidelberg, Germany}

\author[0000-0002-3461-5228]{Ian D. McGreer}
\affil{Steward Observatory, University of Arizona, 933 North Cherry Avenue, Tucson, AZ 85721, USA}

\author[0000-0001-9024-8322]{Bram Venemans}
\affil{Max Planck Institut f\"ur Astronomie, K\"onigstuhl 17, D-69117, Heidelberg, Germany}

\author[0000-0003-4793-7880]{Fabian Walter}
\affil{Max Planck Institut f\"ur Astronomie, K\"onigstuhl 17, D-69117, Heidelberg, Germany}

\author[0000-0002-1318-8343]{Simon Dye}
\affil{School of Physics and Astronomy, Nottingham University, University Park, Nottingham, NG7 2RD, UK}

\author{Brad W. Lyke}
\affil{University of Wyoming, Physics \& Astronomy 1000 E. University, Dept 3905 Laramie, WY 82071, USA}

\author{Adam D. Myers}
\affil{University of Wyoming, Physics \& Astronomy 1000 E. University, Dept 3905 Laramie, WY 82071, USA}

\author[0000-0001-5595-757X]{Evan Haze Nunez}
\affil{Department of Physics and Astronomy, California State Polytechnic University, 3801 West Temple Ave, Pomona, CA 91768, USA}
\affil{Department of Physics and Astronomy, El Camino College, 16007 Crenshaw Blvd, Torrance, CA 90506, USA}

\begin{abstract}
This is the third paper in a series aims at finding reionzation-era quasars with the combination of DESI Legacy imaging Surveys (DELS) and near-infrared imaging surveys, such as the UKIRT Hemisphere Survey (UHS),  as well as the {\it Wide-field Infrared Survey Explore} ({\it WISE}) mid-infrared survey. In this paper, we describe the updated quasar candidate selection procedure, report the discovery of 16 quasars at $6.4\lesssim z \lesssim6.9$ from area of $\sim$13,020 deg$^2$, and present the quasar luminosity function (QLF) at $z\sim6.7$. The measured QLF follows $\Phi(L_{1450})\propto L_{1450}^{-2.35}$ in the magnitude range --27.6$<M_{1450}<$--25.5. We determine the quasar comoving spatial density at $\langle z \rangle$=6.7 and $M_{1450}<-26.0$ to be $\rm 0.39\pm0.11  Gpc^{-3}$ and find that the exponential density evolution parameter to be $k=-0.78\pm0.18$ from $z\sim6$ to $z\sim6.7$, corresponding to a rapid decline by a factor of $\sim 6$ per unit redshift towards earlier epoch, a rate significantly faster than that at $z\sim 3- 5$. The cosmic time between $z\sim6$ and $z\sim6.7$ is only 121 Myrs. The quasar density declined by a factor of more than three within such short time requires that SMBHs must grow rapidly or they are less radiatively efficient at higher redshifts. We measured quasar comoving emissivity at $z\sim6.7$ which indicate that high redshift quasars are highly unlikely to make a significant contribution to hydrogen reionization. The broad absorption line (BAL) quasar fraction at $z\gtrsim6.5$ is measured to be $\gtrsim$22\%. In addition, we also report the discovery of additional five quasars at $z\sim6$ in the appendix.
\end{abstract}


\section{Introduction} \label{intro}
Absorption spectra of $z>6$ quasars reveal complete Gunn-Peterson (GP) absorption troughs, indicating a rapid increase in the intergalactic medium (IGM) neutral fraction towards higher redshift, marking the end of the reionization epoch at $z\sim6$ (see \cite{fan06} and references therein). However, the GP trough saturates even with a small neutral hydrogen fraction of $x _{HI} \gtrsim 10^{-4}$ and becomes insensitive to higher HI densities. If the IGM is mostly neutral, there would be appreciable absorption redward of the wavelength of the Ly$\alpha$ emission line due to the sum of the Rayleigh scattering and would give rise to long wavelength off-resonance absorptions in the form of a damping wing profile\citep[e.g. ][]{me98,madau00}. 

Despite many efforts made in the last decade, there are only three quasars currently known at $z>7$ \citep{mortlock11, banados18, wang18} and $\sim$20 quasars at $z\gtrsim6.5$ \citep[e.g.][]{venemans13,venemans15,matsuoka16,matsuoka17,matsuoka18,mazzucchelli17,reed17,wang17} discovered to date. This is caused by the combination of a rapid decline of quasar spatial number density towards high redshifts \citep[e.g.][]{fan01,mcgreer13,yang16,jiang16} and the lack of deep near-infrared surveys over large (i.e. $\gtrsim$10,000 deg$^2$) sky area.

Currently, the damping wing analyses have been only performed in the line of sight of two known $z>7$ quasars \citep{mortlock11,bolton11,bosman15,greig17,banados18,davies18}, which limits our current knowledge of the reionization history. In spite of that, combining those constraints from quasar absorption spectra with results from the most recent constraints on the declining Ly$\alpha$ visibility\citep[e.g.][]{pentericci14} and the abundance of high-redshift galaxies \citep[e.g.][]{beckwith06,illingworth13} and the CMB polarization measurements\citep{planck18}, current data strongly suggest a peak of reionization activity and emergence of the earliest galaxies and AGNs at $7 < z < 11$ \citep{robertson15}. This highlights the need to expand the search for quasars at $z\gtrsim7$.

Which sources dominate the ionizing photon budget is another key question in understanding the cosmic ionization history \citep[e.g.][]{madau15}. Measurement of quasar luminosity function (QLF) at the epoch of reionization (EoR) directly yields the ionizing radiation output from quasars and will help to solve this longstanding question. In addition, the QLF encodes information about the build-up of SMBHs, provide key insights into understanding the BH growth history \citep[e.g.][]{willott10} and the co-evolution of SMBHs with their hosts \citep[e.g.][]{carilli13,venemans17,decarli18}. 
However, to determine the QLF accurately at high redshift is extremely difficult. Not only does it require a large uniformly selected quasar sample,  but the sample needs to be statistically complete.

The QLF at $z\sim5$ is well measured over a wide luminosity range ($-30\lesssim M_{1450}\lesssim -23$) in the past few years \citep{mcgreer13,yang16,mcgreer18}, which suggests that the QLF can be described with a double power-law function with a very steep bright-end slope of $\beta \sim -3.6$ \citep{yang16}, a flatter faint-end slope of $\alpha \sim -2.0$ \citep{mcgreer13,mcgreer18} and a characteristic magnitude of $M_{1450}^\ast \sim -27$ . However, the parameters of the $z\sim6$ QLF are still debated, \cite{jiang16} found that the bright-end slope is $\beta=-2.8$ using the SDSS quasar sample and changes to $\beta=-2.56$ when including fainter Stripe 82 quasars. The faint-end slope and the characteristic magnitude are even less constrained due to the small number of known faint $z\sim6$ quasars \citep[e.g.;][]{willott10,kashikawa15,jiang16}. More recently, \cite{kulkarni18} found that the QLF at $z\sim6$ has a much steeper bright-end slope ($\alpha = -5.50$), a very bright characteristic magnitude ($M_{1450}^\ast = -29.2$) and a faint-end slope of $\beta=-2.4$. Nevertheless, these results provide one common conclusion: the contribution of quasars to hydrogen reionization at $z\sim5-6$ is subdominant.  

\cite{venemans13} attempted to estimate the quasar luminosity function at $z>6.5$ using three $z>6.5$ quasars over $\sim$ 300 deg$^2$ area from the VISTA Kilo-degree Infrared Galaxy Survey \citep[VIKING;][]{arnaboldi07}, and found that it consistent with the number density of quasars at $z\sim6$ with the exponential density evolution parameter of $k=-0.47$. However their early results have large uncertainties due to the small number of quasars and limited sky coverage. Thus, a large statistically complete and uniformly selected quasar sample at $z>6.5$ is needed to measure the QLF at the EoR.

In \citet[][hereafter Paper I]{wang17}, we demonstrate that the combination of the Dark Energy Spectroscopic Instrument (DESI)\footnote{\url{http://desi.lbl.gov/}} Legacy Imaging Surveys \citep[DELS;][]{dey18}, near-infrared (NIR) surveys like UKIRT Hemisphere Survey \citep[UHS;][]{dye18}, as well as the {\it Wide-field Infrared Survey Explore} \citep[WISE;][]{wright10} mid-infrared survey enables us to search for very high redshift quasars over a large sky coverage. We present the discovery of a $z=6.63$ quasar in Paper I and a luminous $z=7.02$ quasar in \citet[][Paper II, hereafter]{wang18}.

\begin{deluxetable*}{ccccccc}
\tabletypesize{\scriptsize}
\tablecaption{Photometric Information of Datasets Used in This Paper\label{tbl_surveys}}
\tablewidth{0pt}
\tablehead{
\colhead{Survey} & \colhead{Band} & \colhead{Depth (5-$\sigma$) } & \colhead{Sky Area (deg$^{2}$)} & \colhead{AB offset} & \colhead{Ref} 
}
\startdata
DELS (DR4+DR5)\tablenotemark{a}     & $g$, $r$, $z$ & 24.0, 23.5, 22.5 & 13,020 & --- & \cite{dey18}\\
PS1     & $g_{ps1}$, $r_{ps1}$, $i_{ps1}$, $z_{ps1}$, $y_{ps1}$ & 23.3, 23.2, 23.1, 22.3, 21.3 & 30,940 &--- & \cite{chambers16}\\
UHS & $J$ & 20.5& 12,700& 0.938 & \cite{dye18}\\
UKIDSS & $J$ & 20.5& 5,200 & 0.938 & \cite{lawrence07}\\
VHS & $J$ & 21.1& 20,000 & 0.937 &\cite{mcmahon13}\\
VIKING & $J$ & 22.1& 1350 & 0.937 &\cite{arnaboldi07}\\
ALLWISE & $W1$ & 20.3\tablenotemark{b} & 41,253 & 2.699& \cite{wright10}
\enddata
\tablecomments{All magnitudes in this table are in AB system. The conversion factors from VEGA to AB are listed in the fifth column.}
 \tablenotetext{a}{The depth for DELS is for those area with only one photometric pass.}
 \tablenotetext{b}{This value was estimated using the magnitude-error relation from \cite{yang16}, assuming the number of coverage equals to 24, which corresponds to the mean coverage of ALLWISE dataset.}
 
\end{deluxetable*}

In this paper, we present the updated quasar selection procedure by further including data from the Pan-STARRS1 (PS1) Survey \citep{chambers16}. We report the discovery of 16 quasars at $6.4\lesssim z \lesssim 6.9$ as well as 5 quasars at $z\sim6$, and present the QLF at $z\sim6.7$.
The paper is organized as follows: 
in \S\ \ref{sec_selection} we briefly introduce imaging surveys and present our updated quasar selection procedure; 
in \S\ \ref{sec_obs} we describe our follow up spectroscopic observations for quasar candidates; 
in \S\ \ref{sec_results} we report our new discoveries, along with measurements of individual quasar properties, and statistical properties of entire sample;
in \S\ \ref{sec_qlf} we present the selection completeness and derive the $z\sim6.7$ QLF measurement;
in \S\ \ref{sec_discussion} we discuss the quasar spatial density evolution and the contribution of quasars to the cosmic hydrogen reionization.
We summarize our results in \S\ \ref{sec_summary}.
Finally, we report the discovery of five additional $z\sim6$ quasars in the Appendix \ref{sec_app}.

Following our previous papers, optical magnitudes are reported on the AB system with Galactic extinction \citep{SFD98} corrected, near-IR and mid-IR magnitudes are reported on the Vega system. We adopt a standard $\Lambda$CDM cosmology with Hubble constant $H_0=70\,{\rm km~s}^{-1}\,{\rm Mpc}^{-1}$, and density parameters $\Omega_{\rm M}=0.3$ and $\Omega_{\Lambda}=0.7$.

\section{Quasar Candidates Selection} \label{sec_selection}
At $z\sim7$, the $\rm Ly$$\alpha$ emission line in the quasar spectrum redshifts to $\sim1.0\mu$m. Thus, quasars at $z\sim7$ are characterized by their very red $z-J$ color due to the presence of neutral hydrogen at high redshift that absorbs most of the  emission blueward of the $\rm Ly$$\alpha$ emission line in quasar spectra. As a result, we need both deep optical and NIR photometry to select quasars at $z\sim 7$.

\subsection{Imaging Data} \label{sec_imaging}
For optical bands, we mainly used data from DELS\footnote{\url{http://legacysurvey.org/}}, which consists of three different imaging surveys: the Dark Energy Camera Legacy Survey (DECaLS), the Beijing-Arizona Sky Survey \citep[BASS;][]{zou17} and the Mayall $z$-band Legacy Survey (MzLS). These three surveys jointly image $\sim$14,000 deg$^2$ of the extragalactic sky visible from the northern hemisphere in three optical bands ($g$, $r$ and $z$). The DECaLS survey covers $\sim$9000 deg$^2$ extragalactic sky with Decl.$\le32^\circ$ and the BASS+MzLS cover $\sim5000$ deg$^2$ sky with Decl.$\ge32^\circ$. There is a total of $\sim$ 300 deg$^2$ regions at $\rm 32.5^\circ<Decl.<34.5^\circ$ where DECaLS overlaps with BASS$+$MzLS. An overview of DELS surveys can be found in \cite{dey18}. We also include PS1 photometric data in our selection, which provides a 3$\pi$ sky coverage in $g_{ps1}$-, $r_{ps1}$-, $i_{ps1}$-, $z_{ps1}$-, and $y_{ps1}$-bands. The PS1 photometric data was obtained from MAST Casjobs PS1 Archive\footnote{\url{http://mastweb.stsci.edu/ps1casjobs/}}. Although the PS1 survey is shallower than DELS, it has additional $y_{ps1}$-band which is redder and narrower than the reddest $z$-band of DELS. 
At NIR, we combine UHS DR1 with public data from UKIRT Infrared Deep Sky Survey \citep[UKIDSS;][]{lawrence07} DR10, VISTA Hemisphere Survey \citep[VHS;][]{mcmahon13} DR5 and VIKING DR4. The UHS and UKIDSS data were obtained from WFCAM Science Archive\footnote{\url{http://wsa.roe.ac.uk/index.html}}, while the VHS data and VIKING data were obtained from VISTA Science Archive\footnote{\url{http://horus.roe.ac.uk/vsa/}}. In addition, we used the ALLWISE release\footnote{\url{http://wise2.ipac.caltech.edu/docs/release/allwise/}} of the WISE data, which combines  the original WISE survey \citep{wright10} with data from the NEOWISE \citep{mainzer11} post-cryogenic phase. Because UHS DR1 only provides $J$-band photometry, we only used $J$-band in our selection in order to have a homogeneous selection procedure over the whole DELS footprint. The basic characteristics of these imaging surveys are listed in Table \ref{tbl_surveys}.

\subsection{$z\sim7$ Quasar Candidate Selection}\label{zselection}

\begin{figure}
\centering
\includegraphics[width=0.5\textwidth]{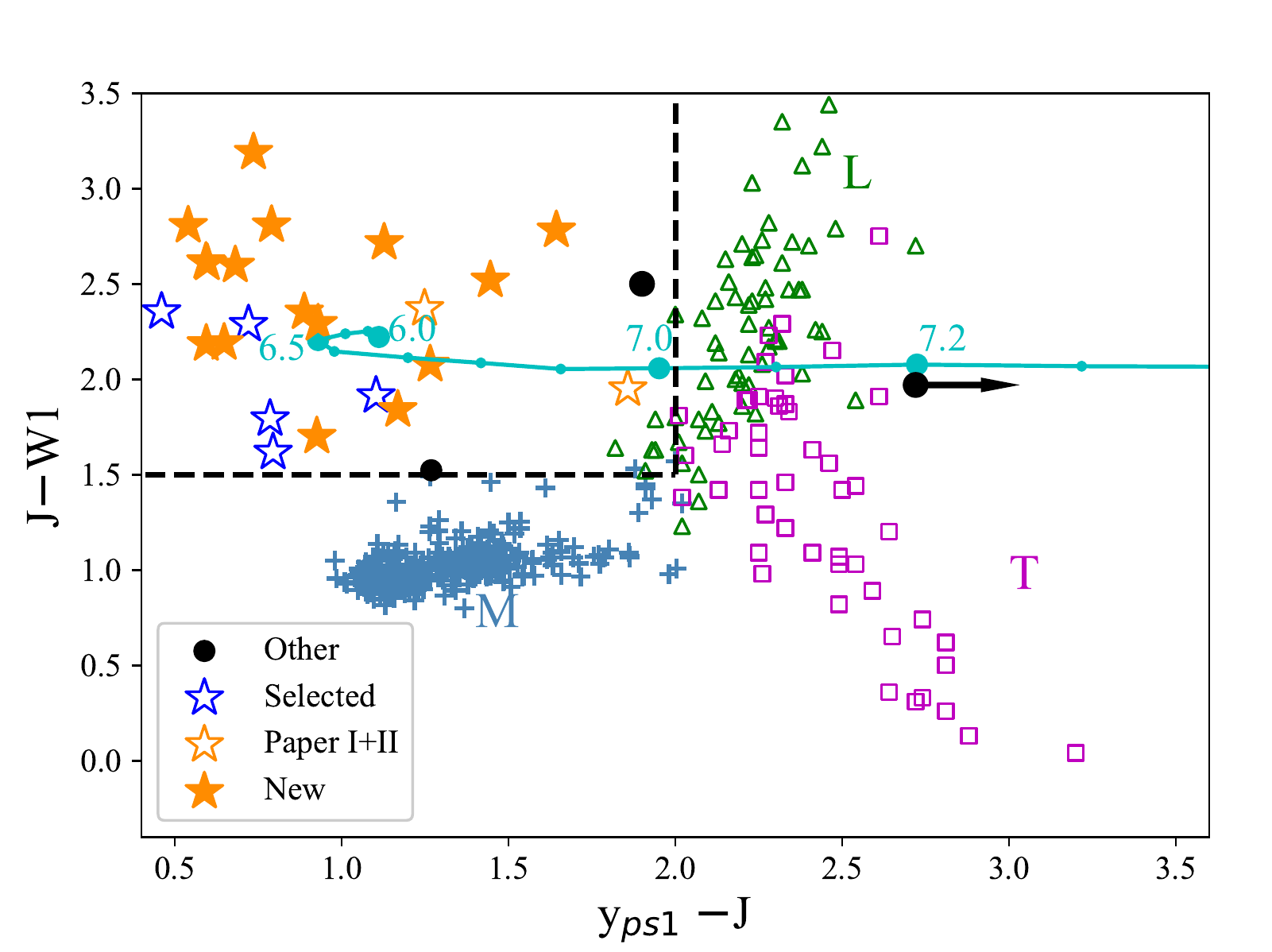}
\caption{The $y_{ps1,AB}-J_{VEGA}$ vs. $J_{VEGA}-W1_{VEGA}$ color-color diagram. The cyan line and cyan filled circles represent the color-redshift relation predicted using simulated quasars \citep{mcgreer13,yang16} from $z = 6.0$ to $z = 7.3$, in steps of $\Delta z = 0.1$. The large cyan circles highlight the colors at $z =$ 6.0, 6.5, 7.0, and 7.2. The orange open asterisks denote two $z>6.5$ quasars reported in Paper I, Paper II and orange solid asterisks denote $z\gtrsim6.4$ quasars found in this paper. The open blue asterisks depict previously known $z>6.5$ quasars that recovered by our selections. The small black circle denotes previously known $z>6.5$ quasar that have all PS1, NIR and WISE detections. The two larger black circles present two known $z>7$ quasars \citep{mortlock11,banados18} with $y_{ps1}$ from forced photometry on PS1 images. The steel blue crosses, green open triangle, and magenta open squares depict the positions of M, L, and T dwarfs, respectively \citep{kirkpatrick11, best15}. 
 \label{fig_color}}
\end{figure}

We started from DELS data release 4 (DR4, MzLS+BASS data) and data release 5 (DR5, DECaLS data), which contains $\sim0.2$ billion and $\sim0.7$ billion sources, respectively. Photometric selection procedure of high-redshift quasar candidates in our survey consists of the following seven steps:
(1) We first select targets which have $z$-band detection but are not observed/not detected (at 5-$\sigma$ level) in $g$ and/or $r$ bands. This results in a total of $\sim0.3$ billion sources. 
(2) Then we cross-matched this catalog with the PS1 data release 1 (DR1) catalog using 2\farcs0 search radius and remove those targets detected in PS1 $g_{ps1}$, $r_{ps1}$ and $i_{ps1}$ bands at 5$\sigma$ levels. 
(3) We then select targets which have both DELS $z$ and PS1 $y_{ps1}$-band with at least 7-$\sigma$ detections, have PS1 $y_{ps1}$-band brighter than 21.5 and fainter than 15 magnitude, and have PS1 $z_{ps1}$-band undetected in 5-$\sigma$ or with PS1 $z_{ps1}-y_{ps1}>1.5$. 
(4) We performed forced photometry on PS1 images using an aperture radius of  a 1\farcs0 for all sources with ALLWISE counterparts within 2\farcs0 separation and removed those with $i_{ps1,forced}$ brighter than 23.1 or $z_{ps1,\rm forced}-y_{ps1,\rm forced}<1.7$. 
(5) We further reject targets identified as extended sources in both PS1 photometry (magnitude differences between aperture photometry and PSF photometry larger than 0.3 ) and DELS photometry (type not equals ``PSF"). Considering the astrometry uncertainties in both PS1 and DELS are relative small (i.e. $\lesssim 0\farcs2$), we further reject targets with distance between PS1 and DELS positions larger than 1\farcs0.
(6) We cross-matched our candidates with infrared photometric catalogs from UHS, UKIDSS, VHS, and VIKING and rejected those targets with $y_{ps1}-J>2.0$ or $J-W1<1.5$ if they are detected in $J$-band. Figure \ref{fig_color} shows the $y_{ps1}-J/J-W1$ color-color diagram of high redshift quasars and Galactic cool dwarfs, as well as our color cuts. 
(7) We finally visually inspecting those candidates. Objects visible in any of PS1 $g_{ps1}$, $r_{ps1}$, $i_{ps1}$ and DELS $g$, $r$ bands, affected by cosmic rays, or contaminated by nearby bright stars are removed.

The selection procedure described above yields a total of  121 quasar candidates left for spectroscopically follow-up  observations. We refer these dropouts as our main quasar candidates in the following sections. Our criteria for targeting $z\sim7$ quasar candidates are summarized as:

\begin{equation}
S/N(g,r,g_{ps1},r_{ps1},i_{ps1})<5.0
\end{equation}
\begin{equation}
S/N(z,y_{ps1})>7.0
\end{equation}
\begin{equation}
z>16, 15<y_{ps1}<21.5
\end{equation}
\begin{equation}
S/N(z_{ps1})<5.0 ~or~ z_{ps1}-y_{ps1}>1.5
\end{equation}
\begin{equation}
i_{ps1,\rm forced}>23.1, z_{ps1,\rm forced}-y_{ps1,\rm forced}>1.7
\end{equation}
\begin{equation}
y_{ps1}-J<2.0
\end{equation}
\begin{equation}
J-W1>1.5
\end{equation}

\subsection{Supplementary Quasar Selection}\label{sselection}
Recently, an ultra-luminous (i.e. $M_{1450}<-29$) quasar population has been discovered at high redshift  \citep{wang15,wu15,wang16}. Such ultra-luminous quasars are detected in the dropout bands due to their extreme brightness. Our selection procedure presented in \S \ref{zselection} requires non-detections in bands bluer than $z$. On the other hand, strong gravitationally lensed quasars would also be missed by requiring non-detections in bluer bands as the lensing galaxy would contribute flux in those bands \citep[e.g.][]{mcgreer10}. In order to recover such quasar population at $z\gtrsim6.5$ and also expand our quasar searching sky area, we started from PS1 DR1 catalog and selected targets with very red colors ($(g_{ps1},r_{ps1},i_{ps1})-y_{ps1}>3.0,~z_{ps1}-y_{ps1}>1.5$) at galactic latitude greater than 5 degrees. Then we cross-matched this catalog with all available NIR photometry data (UHS, VHS, VIKING, 2MASS) and the ALLWISE photometry catalog. We then used the $y_{ps1}-J/J-W1$ color-color diagram (Figure \ref{fig_color}) and $W1-W2>0.4$\citep{wang16} to reject Galactic cool dwarfs. In the work reported here, we only observed eleven high priority candidates in this supplementary  selection for spectroscopic observations. 

\begin{deluxetable*}{lccccccccccccrl}
\tabletypesize{\scriptsize}
\tablecaption{Observational information of 16 new quasars reported in this paper. \label{z7qsoobs}}
\tablewidth{0pt}
\tablehead{
\colhead{Name} & \colhead{Telescope} & \colhead{Instrument} & \colhead{Exposure (sec)} & \colhead{OBS-DATE (UT)} }
\startdata
DELS J041128.63$-$090749.8  & MMT & Red Channel & 2700   & 20171226\\
DELS J070626.39$+$292105.5 & MMT & Red Channel & 2700   & 20171226 \\
DELS J080305.42$+$313834.2 & MMT & Red Channel & 6000 & 20180116, 20180120\\
DELS J082931.97$+$411740.4 & MMT & Red Channel & 2700   & 20171226 \\
DELS J083737.84$+$492900.4 & MMT & Red Channel & 3600   & 20170320 \\
DELS J083946.88$+$390011.5 & MMT & Red Channel & 3600   & 20170321\\
DELS J091013.63$+$165629.8 & MMT & Red Channel & 6300  & 20171226, 20180120\\
DELS J091054.53$-$041406.8 & Magellan/Gemini & FIRE/GMOS-N& 900/4800 & 20180417/20180516\\
DELS J092347.12$+$040254.4\tablenotemark{a} & MMT & Red Channel & 6000   & 20180120, 20180203\\
DELS J110421.59$+$213428.8 & Magellan/LBT & FIRE/MODS\tablenotemark{b} & 300/3600$\times2$ & 20170604/20180112\\
DELS J113508.93$+$501133.0  & MMT & Red Channel & 1800   & 20171226 \\
DELS J121627.58$+$451910.7 & MMT & Red Channel & 2400   & 20180118\\
DELS J131608.14$+$102832.8 & Magellan/MMT & FIRE/Red Channel & 500/2700 & 20170604/20171226\\
DELS J153532.87$+$194320.1  & Magellan/LBT & FIRE/MODS & 600/4500 & 20170604/20180707 \\
DELS J162911.29$+$240739.6\tablenotemark{c} &  Hale 200 inch & DBSP & 15900 & 20160908, 20160909, 20160911\\
VHS J210219.22$-$145854.0  & Magellan/KECK & FIRE/DEIMOS & 900/12000 & 20170801/20170914
\enddata
 \tablenotetext{a}{This quasar was discovered by \cite{matsuoka18} independently.}
 \tablenotetext{b}{We used binocular mode by using both MODS1 and MODS2.}
 \tablenotetext{c}{This quasar was discovered by \cite{mazzucchelli17} independently.}
\end{deluxetable*}

\begin{figure*}[tb]
\centering
\includegraphics[width=1.0\textwidth]{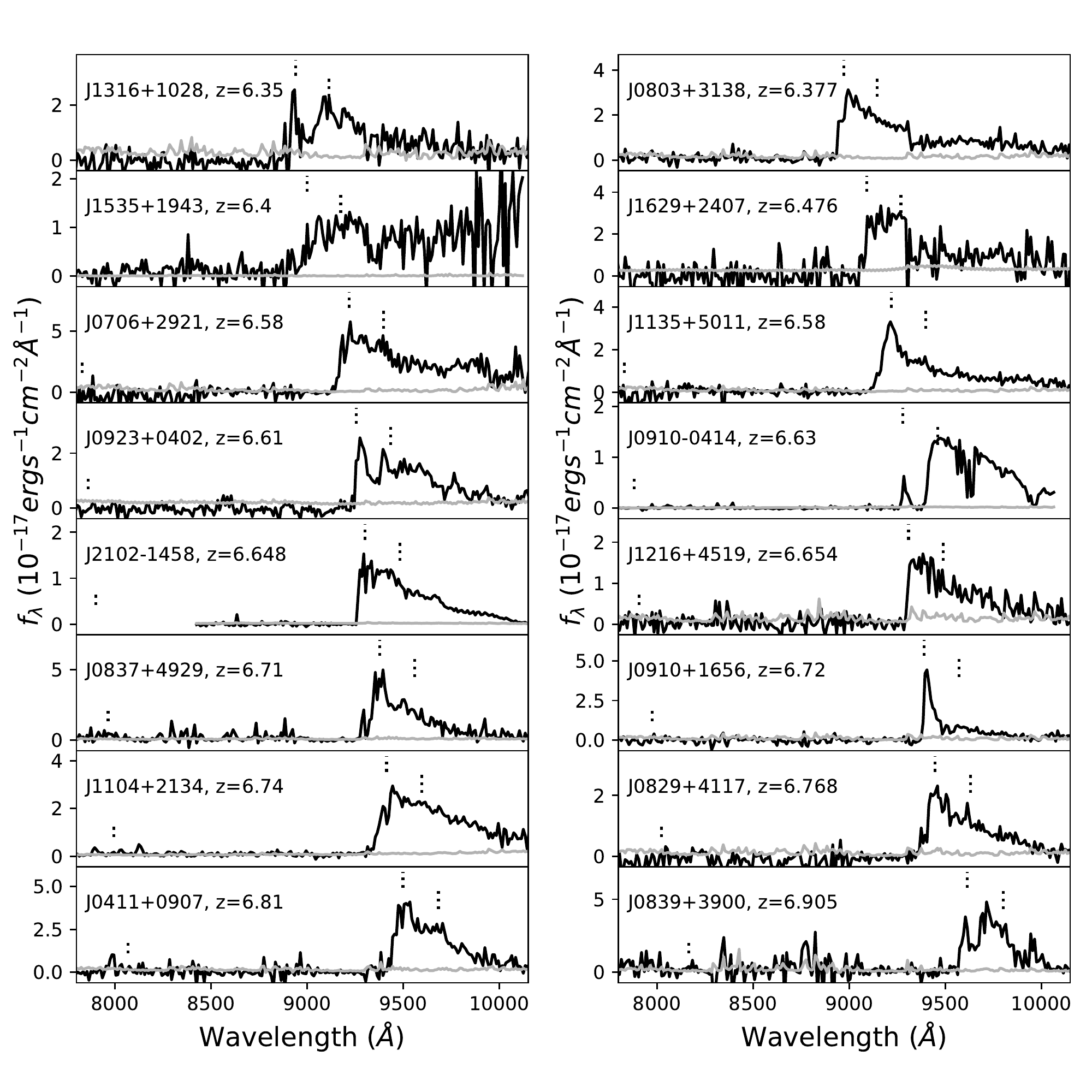}
\caption{Spectra of the 16 newly discovered $z\gtrsim 6.4$ quasars. The redshifts of these newly discovered quasars are list in Table \ref{newz7qso}. The dotted lines mark the positions of three emission lines: $\rm Ly\beta$, $\rm Ly\alpha$, and N\,{\sc v} from left to right. The grey lines denote 1$\sigma$ flux error vector. All spectra are binned into 10 \AA\ in wavelength space using 1/$\sigma^2$ weighted mean algorithm.
\label{z7spec}}
\end{figure*}

\section{Spectroscopic Observations} \label{sec_obs}

We obtained spectroscopic follow-up observations of the quasar candidates with the MMT/Red Channel spectrograph \citep{schmidt89}, MMT/MMIRS  \citep{mcleod12}, MMT/ Binospec  \citep{fabricant98}, Magellan/FIRE \citep{simcoe08}, Magellan/LDSS3-C \citep{stevenson16}, KECK/DEIMOS \citep{faber03}, LBT/MODS \citep{byard00}, Gemini/GMOS \citep{hook04} and P200/DBSP \citep{oke82}, over $\sim$25 observing nights from April 2016 to July 2018.

We observed 33 candidates with MMT 6.5 m telescope, with 26 main candidates targeted with Red Channel Spectrograph, 5 main candidates targeted with MMIRS spectrograph, one main candidate and one supplementary candidate observed with Binospec. We used the 270 l mm$^{-1}$ grating on Red Channel spectrograph centered at 9000 \AA , providing wavelength coverage from 7200 \AA\ to 10800 \AA . We used the $1\farcs0$ or $1\farcs5$ slits depends on seeing conditions, providing resolutions of $R\sim640$ and $R\sim510$, respectively. MMIRS is a wide-field near-IR imager and multi-object spectrograph. We used the HK grism with zJ filter, which provides a wavelength coverage from 9500 \AA\ to 1.3 $\mu$m and a resolution of $R\sim1,000$ with a 1\farcs0 slit. The Binospec is a new imaging spectrograph with dual $8'\times15'$ fields of view. We used 270 grating centered at 7400 \AA\ with a 1\farcs0 slit, which provide a resolution of $R\sim1340$ and wavelength coverage from 6130 \AA\ to 10120 \AA .

We observed 27 main candidates and 7 supplementary candidates with Magellan/FIRE. FIRE is an IR echelle/longslit spectrograph on the 6.5 m Magellan/Baade Telescope. In order to improve the efficiency, we used the high throughput mode which provides a resolution of $R\sim300$--500 from K-band to J-band. The typical exposure time for each target is 5--10 minutes. We further observed those FIRE confirmed high-redshift quasars with MMT/Red Channel, LBT/MODS, KECK/DEIMOS and Gemini/GMOS-N to get optical spectra. 

We observed two main candidates with DBSP on Palomar 200 inch Hale telescope. We used the G316 grating centered at 7500 \AA\ with a 1\farcs5 slit, which provides a resolution of $R\sim 1000$ at 7500 \AA . Two main candidates were targeted with LBT/MODS using the red grating and a 1\farcs2 slit, which delivers a resolution of $R\sim 1200$ from 5000 \AA\ to 1 $\mu$m. Two main candidates were observed with Magellan/LDSS3-C using the VPH red grating with 1\farcs0 slit, which delivers a resolution of $R\sim 1360$ from 6000 \AA\ to 1 $\mu$m. We observed three supplementary candidates using KECK/DEIMOS with 830G grating.

In total, we spectroscopically observed 65 main candidates and 11 supplementary candidates. The data obtained from Red Channel, Binospec, MODS and DBSP were reduced using standard IRAF routines. The GMOS data were reduced using the Gemini IRAF packages. The data obtained with FIRE and MMIRS were reduced using a custom set of Python routines \citep{wang17} which includes dark subtraction, flat fielding, sky subtraction, wavelength calibration and flux calibration. The data obtained with DEIMOS were reduced using the XIDL\footnote{\url{http://www.ucolick.org/~xavier/IDL/}} suite of astronomical routines in the Interactive Data Language (IDL), which was developed by X. Prochaska and J. Hennawi. 

\section{Results} \label{sec_results}
\subsection{Discovery of 16 New Quasars at $6.4\lesssim z\lesssim6.9$} \label{discoveries}
Four spectroscopic observed candidates do not have any signal due to poor weather conditions. Other spectroscopic observed candidates were identified to be high redshift quasars, Galactic cool dwarfs or ``Non-Quasars" without any obvious break and/or emission line which are the main features of high redshift quasars\footnote{The list of non-quasars is available from the corresponding author on reasonable request.}. In Paper I and Paper II, we have reported two quasars from the main sample, DELS J1048--0109 at $z=6.6759$ and DELS J0038--1527 at $z=7.02$. In \cite{fan18}, we reported one strong gravitationally lensed quasar (UHS J0439+1634) at $z=6.511$ from the supplementary sample, including followup {\em HST} imaging and detailed lens model results. All redshifts measured from (sub)mm emission lines from host galaxy have four digits, redshifts measured from \Mgii\ broad emission lines have three digits, and redshifts measured from Ly$\alpha$ have two digits throughout the paper. 

Here, we report additional 16 new quasars at $6.4\lesssim z\lesssim 7.0$. Fifteen of them come from our main sample, and one object, J210219.22--145854.0, is from our supplementary sample. J2102--1458 is not covered by our main selection because it is outside the DELS footprint and thus not included in our QLF measurements in the following sections.  The details of the spectroscopic observations of these newly discovered quasars are listed in Table \ref{z7qsoobs}. Figure \ref{z7spec} shows the discovery spectra of these quasars. For those quasars with spectra available only in optical, we measured quasar redshifts from $\rm Ly\alpha$ and \Nv\ lines by fitting the observed spectra to the SDSS quasar template \citep{vanden01} using a visual recognition assistant for quasar spectra software \citep[ASERA;][]{yuan13}. The typical redshift errors are about 0.03 due to the combination of low spectral resolution and strong absorptions blueward of $\rm Ly\alpha$. 
We have already obtained near-IR spectra for some of these quasars, for which we estimate the redshifts by fitting \Mgii\ emission lines (Yang J. {\it in preparation}). The redshift uncertainties for quasars that have near-IR spectra are usually around 0.01. Note that the redshift uncertainties quoted here do not take the possible shift compared with (sub)-millimeter emission lines from quasar host galaxies, which could be up to several thousand kilo-meter per seconds \citep[e.g.][]{decarli18}. 
We listed redshifts and photometric information of newly discovered quasars in Table \ref{newz7qso}. There are four quasars that do not have any available NIR photometry. For these objects, we obtained additional $J$-band photometry with UKIRT/WFCAM \citep{casali07}. 

We use the continuum magnitude at rest-frame 1450 \AA\ in the determination of the QLF. At $z>6.3$, the rest-frame 1450 \AA\ is red shifted to observed wavelengths longer than 1.06 $\mu$m, which is beyond the useful wavelength coverage of our optical spectra. Thus it is not possible to estimate the 1450 \AA\ magnitudes by directly fitting power-law to the discovery spectra shown in Figure \ref{z7spec}. Instead, we scale the composite spectra of luminous low redshift quasars \citep{selsing16} to the Galactic extinction corrected J-band photometry of each quasar. Then we estimate the 1450 \AA\ magnitudes from the scaled composite spectrum. In table \ref{newz7qso}, we list the apparent and absolute AB magnitudes at rest-frame 1450 \AA\ in Column 3 ($m_{1450}$) and Column 4 ($M_{1450}$), respectively.

\begin{deluxetable*}{lccccccccccccrl}
\tabletypesize{\scriptsize}
\tablecaption{Properties of 16 new $z\gtrsim6.4$ quasars reported in this paper. \label{newz7qso}}
\tablewidth{0pt}
\tablehead{
\colhead{Name} & \colhead{Redshift} & \colhead{$m_{1450}$} & \colhead{$M_{1450}$} &
\colhead{$z_{\rm DELS, AB}$} & \colhead{$y_{\rm ps1, AB}$} & \colhead{$J_{\rm VEGA}$} & \colhead{$W1_{\rm VEGA}$} & \colhead{NIR Survey} & \colhead{QLF}
}
\startdata
DELS J083946.88$+$390011.5 & 6.905$\pm$0.01\tablenotemark{c} & 20.63$\pm$0.20 & -26.29$\pm$0.20 & 20.92$\pm$ 0.04 & 20.24$\pm$ 0.08 & 19.45$\pm$ 0.20 & 16.64$\pm$ 0.09 & UHS & Y\\
DELS J041128.63$-$090749.8 & 6.81$\pm$0.03 & 20.28$\pm$0.12 & -26.61$\pm$0.12 & 20.68$\pm$ 0.03 & 20.02$\pm$ 0.06 & 19.13$\pm$ 0.11 & 16.78$\pm$ 0.09 & New\tablenotemark{d} & Y\\
DELS J082931.97$+$411740.4 & 6.768$\pm$0.006\tablenotemark{c} & 20.52$\pm$0.15 & -26.36$\pm$0.15 & 21.36$\pm$ 0.04 & 20.61$\pm$ 0.11 & 19.34$\pm$ 0.15 & 17.26$\pm$ 0.14 & UHS & Y\\
DELS J110421.59$+$213428.8 & 6.74$\pm$0.04 & 20.21$\pm$0.13 & -26.67$\pm$0.13 & 21.06$\pm$ 0.03 & 19.94$\pm$ 0.06 & 19.01$\pm$ 0.12 & 17.31$\pm$ 0.16 & UHS & Y\\
DELS J091013.63$+$165629.8 & 6.72$\pm$0.03 & 21.30$\pm$0.14 & -25.57$\pm$0.14 & 22.09$\pm$ 0.13 & 20.80$\pm$ 0.15 & 20.12$\pm$ 0.13 & 17.52$\pm$ 0.20 & New\tablenotemark{d} & Y\\
DELS J083737.84$+$492900.4 & 6.710$\pm$0.008\tablenotemark{c} & 20.45$\pm$0.18 & -26.42$\pm$0.18 & 20.66$\pm$ 0.02 & 19.86$\pm$ 0.06 & 19.27$\pm$ 0.17 & 17.08$\pm$ 0.11 & UHS & Y\\
DELS J121627.58$+$451910.7 & 6.654$\pm$0.01\tablenotemark{c} & 21.27$\pm$0.14 & -25.58$\pm$0.14 & 21.78$\pm$ 0.12 & 20.62$\pm$ 0.09 & 20.08$\pm$ 0.13 & 17.27$\pm$ 0.13 & New\tablenotemark{d} & Y\\
VHS J210219.22$-$145854.0 & 6.648$\pm$0.01\tablenotemark{c} & 21.36$\pm$0.20 & -25.50$\pm$0.20 &      --- & 20.80$\pm$ 0.14 & 20.21$\pm$ 0.20 & 17.59$\pm$ 0.22 & New\tablenotemark{d} & N\\
DELS J091054.53$-$041406.8 & 6.63$\pm$0.03 & 20.49$\pm$0.15 & -26.36$\pm$0.15 & 21.85$\pm$ 0.14 & 20.76$\pm$ 0.13 & 19.31$\pm$ 0.14 & 16.79$\pm$ 0.10 & VHS & Y\\
DELS J092347.12$+$040254.4\tablenotemark{a} & 6.61$\pm$0.03 & 20.23$\pm$0.11 & -26.61$\pm$0.11 & 21.18$\pm$ 0.02 & 20.20$\pm$ 0.08 & 19.08$\pm$ 0.09 & 16.36$\pm$ 0.07 & LAS & Y\\
DELS J070626.39$+$292105.5 & 6.58$\pm$0.03 & 19.33$\pm$0.08 & -27.51$\pm$0.08 & 20.02$\pm$ 0.02 & 19.15$\pm$ 0.03 & 18.22$\pm$ 0.05 & 15.93$\pm$ 0.06 & UHS & Y\\
DELS J113508.93$+$501133.0 & 6.58$\pm$0.03 & 20.65$\pm$0.17 & -26.19$\pm$0.17 & 20.62$\pm$ 0.04 & 20.12$\pm$ 0.07 & 19.47$\pm$ 0.16 & 17.28$\pm$ 0.12 & UHS & Y\\
DELS J162911.29$+$240739.6\tablenotemark{b} & 6.476$\pm$0.004\tablenotemark{c} & 20.50$\pm$0.18 & -26.32$\pm$0.18 & 20.75$\pm$ 0.03 & 20.00$\pm$ 0.06 & 19.40$\pm$ 0.17 & 16.79$\pm$ 0.08 & UHS & Y\\
DELS J153532.87$+$194320.1 & 6.4$\pm$0.05 & 19.79$\pm$0.13 & -27.01$\pm$0.13 & 20.74$\pm$ 0.04 & 20.35$\pm$ 0.06 & 18.70$\pm$ 0.11 & 15.92$\pm$ 0.05 & UHS & N\\
DELS J080305.42$+$313834.2 & 6.377$\pm$0.006\tablenotemark{c} & 20.28$\pm$0.14 & -26.51$\pm$0.14 & 20.69$\pm$ 0.10 & 20.35$\pm$ 0.10 & 19.18$\pm$ 0.12 & 17.34$\pm$ 0.16 & UHS & N\\
DELS J131608.14$+$102832.8 & 6.35$\pm$0.04 & 21.06$\pm$0.17 & -25.73$\pm$0.17 & 21.39$\pm$ 0.04 & 20.68$\pm$ 0.10 & 19.94$\pm$ 0.15 & 16.75$\pm$ 0.09 & LAS & N\\
\enddata
\tablenotetext{}{{\bf Note:} The column ``NIR Survey" indicate where the $J$ band photometry comes from. The last column indicates whether a quasar is used for QLF measurements.} 
 \tablenotetext{a}{This quasar was discovered by \cite{matsuoka18} independently.}
 \tablenotetext{b}{This quasar was discovered by \cite{mazzucchelli17} independently.}
 \tablenotetext{c}{The redshift measured from near-infrared spectra by fitting broad \Mgii\ emission lines (Yang, J., et al. 2018 {\it in preparation}). }
 \tablenotetext{d}{The J-band photometry were obtained using UKIRT/WFCam (Programs: U/17B/UA01 and U/17B/D04).}
\end{deluxetable*}

\begin{figure}
\centering
\includegraphics[width=0.5\textwidth]{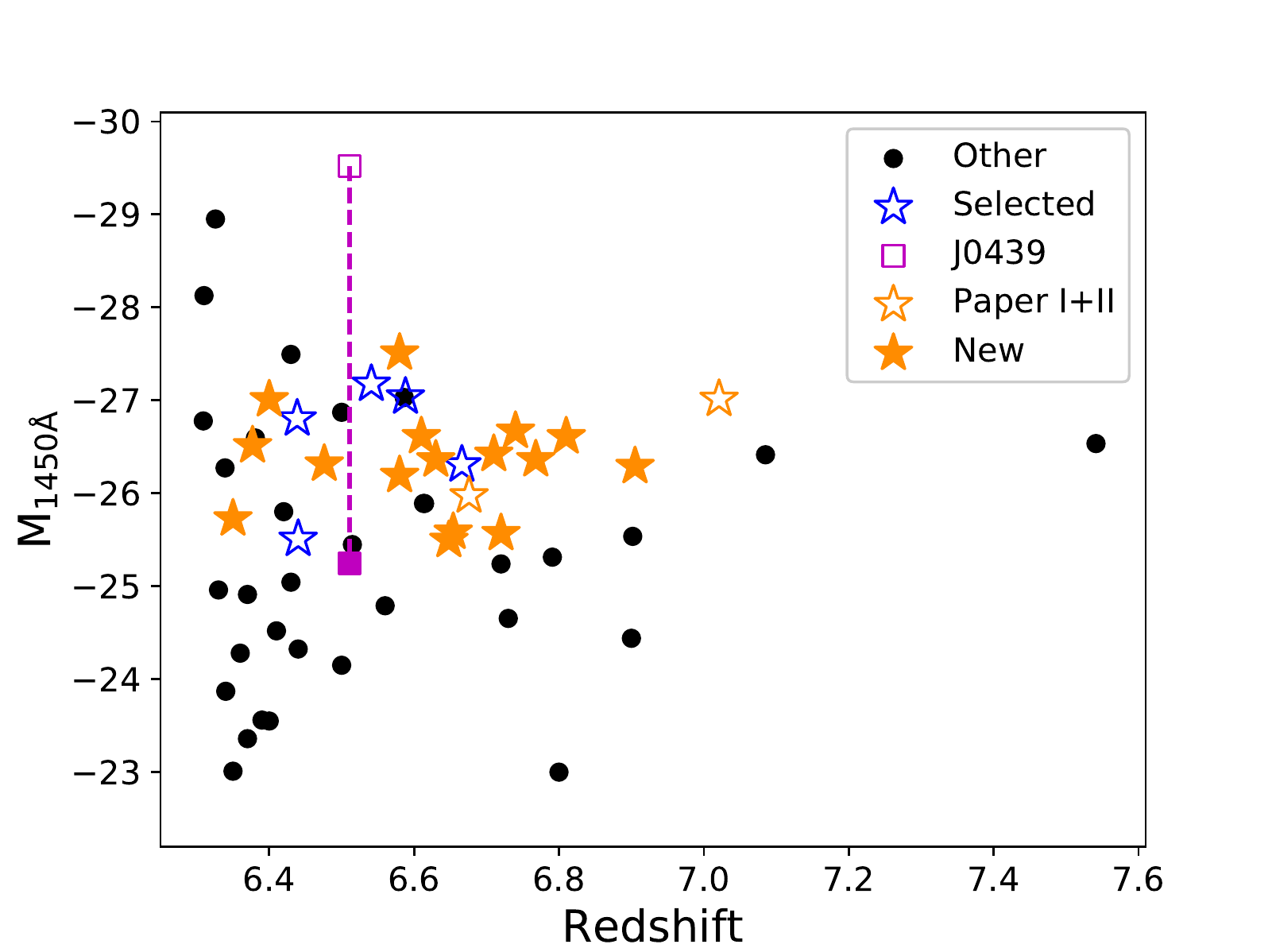}
\caption{The redshift and absolute magnitude distribution of $z\ge$6.3 quasars. The black circles denote previously known quasars missed by our selection because of their faintness, being outside our searching footprint or at relative low redshift and therefore do not satisfy our dropout selection. The two $z>7$ known quasars also not selected by us due to they are un-detected by PS1 survey. The blue open asterisks represent previously known quasars that satisfy our selection procedures and the orange open asterisks denote two quasars published in our Paper I and Paper II. The solid asterisks denote our newly discovered quasars reported in this paper. The magenta squares denote the most distant gravitationally lensed quasar, J0439+1634 at $z=6.511$, before (open) and after (solid) lensing correction \citep{fan18}. \label{mz}}
\end{figure}

Figure \ref{mz} shows the redshift and $M_{1450}$ distribution of all quasars at $z\ge6.3$ from the literature as well as our newly discovered quasars. The on-going Subaru High-z Exploration of Low-Luminosity Quasars (SHELLQs) project \citep{matsuoka16,matsuoka17,matsuoka18} are focusing on faint quasars (i.e. $M_{1450}\gtrsim-25.5$). In comparison, we are focusing on  the parameter space on the redshift and magnitude plane (i.e. at $z>6.5$ and $M_{1450}<-25.5$), where only $\sim$10 previously known quasars were discovered from multiple quasar surveys \citep{mortlock11,venemans13,venemans15,tang17,mazzucchelli17,reed17,banados18}. More importantly, these newly discovered bright $z>6.5$ quasars are crucial for probing the cosmic reionization history by searching for and investigating the damping wing absorption features with future high quality spectroscopy and more accurate redshift measurements.

\subsection{Notes on Individual Quasars}

{\bf DELS J083946.88+390011.5}. J0839+3900 is a broad absorption line (BAL) quasar at $z=6.905$ with strong blue-shifted \Nv\ absorption. We obtained deep Gemini/GNIRS spectrum which shows that J0839+3900 is a Low-ionization BAL (LoBAL) quasar with strong blue shifted \Mgii\ absorptions. It is the highest redshift known LoBAL quasar. The NIR spectrum and related physical parameter measurements will be reported in the future together with NIR spectra of other $z>6.5$ quasars in a subsequent paper (Yang, J. {\it in preparation}). 

{\bf DELS J092347.12+040254.4}. J0923+0402 is a BAL quasar. It was independently discovered by \cite{matsuoka18}.  

{\bf DELS J091054.53--041406.8}. J0910--0414 is a BAL quasar. It was initially identified as a $z\sim6.8$ quasar from our low SNR FIRE spectrum. However, after obtaining a deep GMOS optical spectrum, we found that it is a $z=6.63$ BAL quasar where most of the flux blueward of \Nv\  is absorbed.

{\bf DELS J070626.39+292105.5}. J0706+2921 is the most luminous $z>6.5$ quasar  known to date (J0439+1634 reported in \citep{fan18} is brighter but has lower intrinsic luminosity after correcting for lensing magnification). It has an absolute magnitude of $M_{1450}$=--27.51, and is about 0.3 magnitude brighter than the previous record holder \citep{venemans15}. 

{\bf DELS J162911.29+240739.6}. J1629+2407 was independently discovered and reported by \cite{mazzucchelli17}.

{\bf DELS J153532.87$+$194320.1}. J1535+1943 has a very red $y_{ps1}-J$ color of $\sim$1.7 compared with other quasars ($\sim 1.0$) at similar redshifts. The low SNR spectrum shown in Figure \ref{z7spec} shows that the break at blueward of Ly$\alpha$ is not as sharp as others, which suggests J1535+1943 might be a reddened quasar or has a proximate damped Lyman-alpha (PDLA) system in front of it. 

{\bf DELS J131608.14+102832.8}. J1316+1028 is a BAL quasar at $z=6.35$ with strong blue-shifted \Nv\ absorption. 

\subsection{BAL Quasar Fraction}
BAL quasars show  gaseous outflows which cause strong blue-shifted absorptions in quasar spectra. Previous studies based on spectral analyses indicated that observed BAL quasars comprise about $\sim$15\% of the quasar population at low and intermediate redshifts, without significant redshift dependence \citep[e.g.][]{reichard03,hewett03,knigge08,gibson09}. However, \cite{allen11} found a strong redshift dependence of the BAL quasar fraction with a factor of $3.5\pm 0.4$ decrease from $ z\sim 4.0$, down to $z\sim2.0$. The redshift dependence implies that orientation effect alone is not sufficient to explain this trend. 

An alternative model which allows cosmic evolution of the BAL quasar fraction is that radiation driven winds are the likely origin of quasar outflows \citep[e.g.][]{risaliti10}. The radiation driven winds can be generated by the quasar accretion disc under a variety of physical conditions, and the possibility of their existence is a function of physical parameters such as the BH mass, Eddington ratio, and X-ray to UV flux ratio \citep[e.g.][]{risaliti10}. Thus, investigating whether the BAL quasar fraction is different at the EoR would help us understand the nature of BAL quasars and probe whether it is only related to the orientation or also affected by other physical parameters. 

Four newly discovered quasars in our sample that show strong blue-shifted \Nv\ absorptions, indicating that they are BAL quasars, J1316+1028 at $z=6.35$, J0923+0402 at $z=6.61$, J0910--0414 at $z=6.63$, and J0839+3900 at $z=6.905$. In addition, the NIR spectrum of J0038--1527 shows that it has multiple strong \Civ\ broad absorption troughs (Paper II).  We only estimate the BAL fraction in our main quasar sample because it is statistically complete. There are five published PS1 quasars that also satisfy our main selection procedure (See Table \ref{samplez7qso}). Thus, we need to include these five PS1 quasars, J1048--0109 reported in Paper I and J0038--1527 reported in Paper II, but exclude J2102--1458 when counting the BAL fraction. We visually inspected the spectra of five PS1 quasars \citep{venemans15, mazzucchelli17}, only PSO J036+03 shows small possible absorption troughs at the blueward of \Siv\ and \Civ\ emission lines. However, the spectrum of PSO J036+03 in \cite{venemans15} shows several unusual bumps, which could be due to flux calibration issues typical of Echelle spectra. We treat PSO J036+03 as a possible BAL quasar. Therefore, the observed BAL quasar faction in our main quasar sample is 5(6)/23=21.7(26.1)\%, which is slightly higher than that at lower redshift \citep[e.g.][]{reichard03,hewett03,knigge08,gibson09}. Note that the spectra of some quasars do not cover the \Civ\ lines and we do not know whether they are real Non-BAL quasars. In addition, the color selection bias would underestimate the BAL quasar fraction by a few percents \citep{reichard03} due to that BAL quasars usually have slightly redder colors. Thus, the observed BAL quasar fraction given here should be treated as a lower limit. We will revisit this question in details after collecting NIR spectra for all quasars.

\begin{deluxetable*}{cccccccccccccrl}
\tabletypesize{\scriptsize}
\tablecaption{Previously known $z\gtrsim6.4$ quasars recovered by our main selection procedure. \label{samplez7qso}}
\tablewidth{0pt}
\tablehead{
\colhead{Name} & \colhead{Redshift} & \colhead{$m_{1450}$} & \colhead{$M_{1450}$} &
\colhead{$z_{\rm DELS, AB}$} & \colhead{$y_{\rm ps1, AB}$} & \colhead{$J_{\rm VEGA}$} & \colhead{$W1_{\rm VEGA}$} & \colhead{Ref.\tablenotemark{a}} & \colhead{QLF}
}
\startdata
DELS J003836.10$-$152723.6 & 7.021$\pm$0.006 & 19.93$\pm$0.08 & -27.01$\pm$0.08 & 21.65$\pm$ 0.08 & 20.61$\pm$ 0.10 & 18.75$\pm$ 0.07 & 16.80$\pm$ 0.10 & (1) & Y\\
DELS J104819.08$-$010940.4 & 6.6759$\pm$0.0005 & 20.89$\pm$0.18 & -25.97$\pm$0.18 & 21.95$\pm$ 0.05 & 20.96$\pm$ 0.14 & 19.71$\pm$ 0.17 & 17.34$\pm$ 0.17 & (2) & Y\\
PSO J338.2298$+$29.5089 & 6.666$\pm$0.0004 & 20.55$\pm$0.15 & -26.31$\pm$0.15 & 21.05$\pm$ 0.05 & 20.22$\pm$ 0.09 & 19.43$\pm$ 0.14 & 17.81$\pm$ 0.21 & (3) & Y\\
PSO J323.1382$+$12.2986 & 6.5881$\pm$0.0003 & 19.80$\pm$0.12 & --27.04$\pm$0.12 & 19.72$\pm$ 0.02 & 19.17$\pm$ 0.02 & 18.71$\pm$ 0.11 & 16.36$\pm$ 0.07 & (4) & Y\\
PSO J036.5078$+$03.0498 & 6.541$\pm$0.002 & 19.66$\pm$0.12 & -27.18$\pm$0.12 & 19.98$\pm$ 0.01 & 19.30$\pm$ 0.03 & 18.52$\pm$ 0.10 & 16.73$\pm$ 0.08 & (3) & Y\\
PSO J261.0364$+$19.0286 & 6.44$\pm$0.05 & 21.30$\pm$0.19 & -25.51$\pm$0.19 & 21.60$\pm$ 0.08 & 20.92$\pm$ 0.12 & 20.20$\pm$ 0.18 & 17.91$\pm$ 0.21 & (4) & N\\
PSO J183.1124$+$05.0926 & 6.4386$\pm$0.0004 & 20.00$\pm$0.10 & -26.80$\pm$0.10 & 20.53$\pm$ 0.01 & 19.98$\pm$ 0.05 & 18.88$\pm$ 0.08 & 16.97$\pm$ 0.13 & (4) & N\\
\enddata
\tablenotetext{a}{These objects were discovered by several studies: (1) Paper II; (2) Paper I; (3) \cite{venemans15}; (4) \cite{mazzucchelli17}.}
\end{deluxetable*}

\subsection{Radio Properties}
\begin{figure}
\centering
\includegraphics[width=0.5\textwidth]{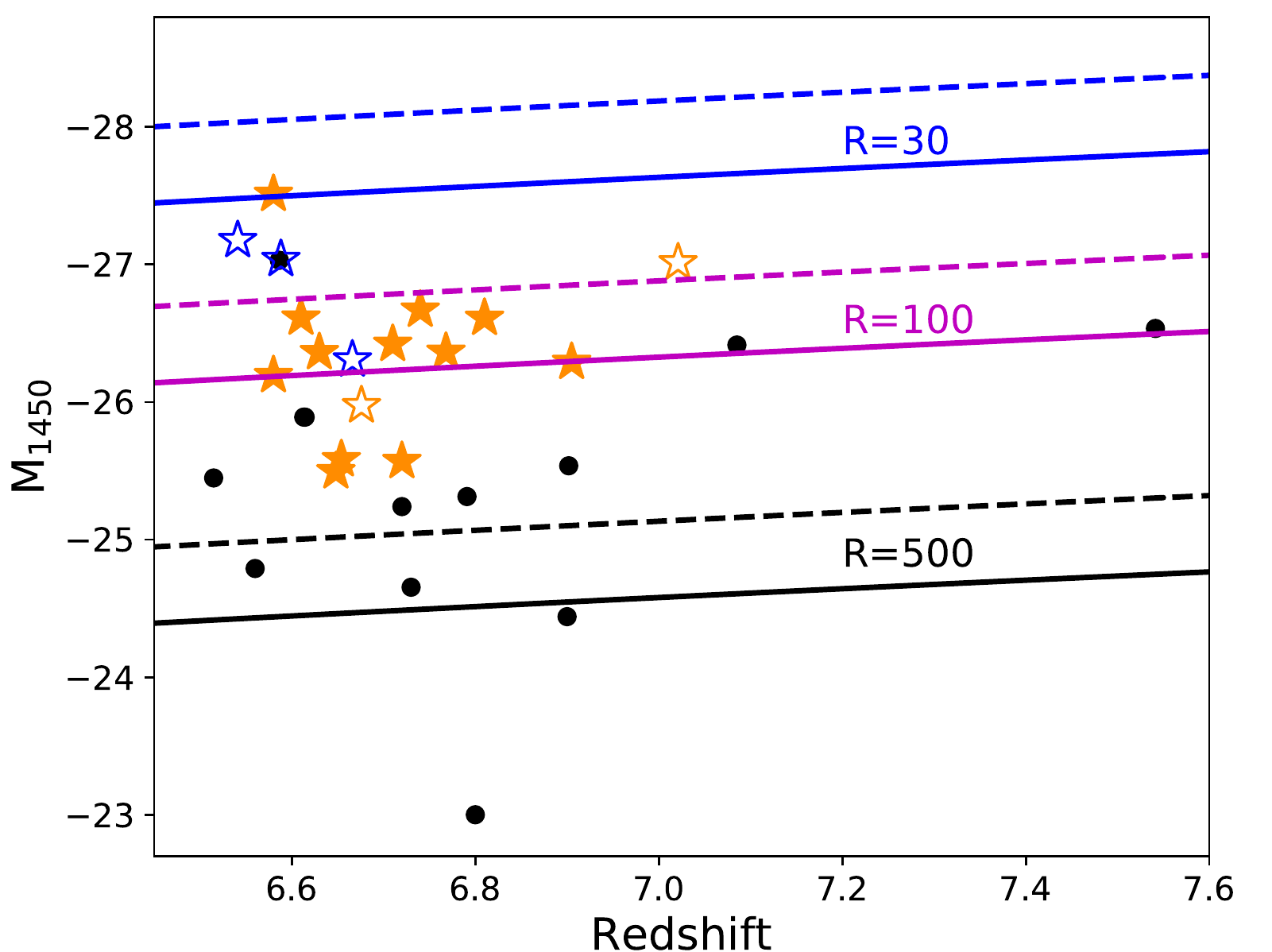}
\caption{The redshift and absolute magnitude distribution and radio loudness constraints of $z>$6.5 quasars. The symbols are the same as in Figure \ref{mz}. The solid lines denote radio loudness calculated from FIRST flux limit and dashed lines denote  radio loudness calculated from NVSS flux limit. The black, magenta and blue curves represent $R=500$, $R=100$, and $R=30$, respectively. It shows that both FIRST and NVSS surveys are too shallow to rule out those $z>6.5$ quasars are radio loud quasars with $R\sim 100$. But none of our newly discovered quasars is extremely radio loud quasar with R$>$500.
\label{fig_radio}}
\end{figure}

\begin{figure}
\centering
\includegraphics[width=0.5\textwidth]{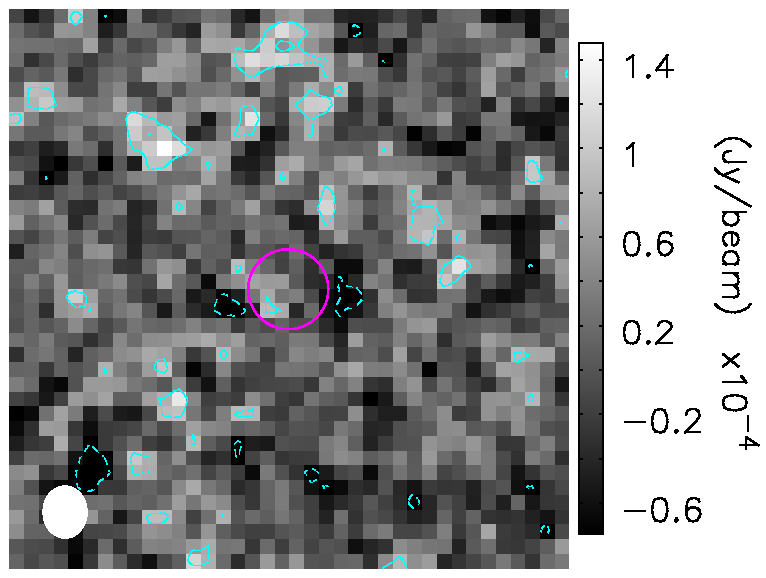}
\caption{Stacked VLA FIRST image of 17 $z>6.5$ quasars. The white ellipse shows the beam size and the magenta circle (radius of 5 arcsec) marks the position of stacked quasar position. The pixel size of the stacked image is 1.8 arcsec/pixel. The rms of the stacked image is $\sim$3.7$\times$10$^{-2}$ mJy/beam, four times better than FIRST depth. +2$\sigma$ contours are shown as cyan solid lines and --2$\sigma$ contours are shown as cyan dashed lines. No signal is detected at the quasar position.
\label{fig_stack}}
\end{figure}

Bright radio sources at the EoR allow us to study the cosmic reionization by detecting 21 cm absorptions from intervening neutral IGM \citep[e.g.][]{carilli07, semelin16}. In addition, powerful radio jets play a key role in the formation and build-up of SMBHs \citep[e.g.][]{volonteri15}. The radio loud fraction of quasars is found to be $\sim$10\% from low redshifts up to $z\sim6$ \citep[e.g.][]{jiang07,banados15}. However, no radio loud quasar has been found at $z>6.5$, where the universe is relatively neutral. The large quasar sample presented here allows investigation of the radio properties of early quasars. We cross-matched all known $z\gtrsim6.5$ quasars presented in this paper and from literatures with the Faint Images of the Radio Sky at Twenty Centimeters \citep[FIRST;][]{becker95} survey, and the  NRAO VLA Sky Survey \citep[NVSS;][]{condon98}. However, none of the known $z>6.5$ quasars is detected in these two radio surveys. 

In order to distinguish radio loud quasars from radio quiet quasars, many criteria were proposed in the literatures \citep[e.g.][]{kellermann89, stocke92}. Here, we calculate the radio loudness as $R=f_{6\rm cm}/f_{2500}$ \citep[e.g.][]{stocke92,jiang07}, where $f_{6\rm cm}$ and $f_{2500}$ are flux densities at rest-fame 6cm and 2500 \AA\, respectively. We estimate the $f_{2500}$ from $m_{1450}$ by assuming $f_\nu \propto \nu^{-0.6}$ \citep{lusso15}, and estimate $f_{6\rm cm}$ from the 1.4GHz observed flux density by assuming $f_\nu \propto \nu^{-0.75}$ \citep[e.g.][]{wang07}. Figure \ref{fig_radio} shows the constraints on radio loudness of $z>6.5$ quasars with the flux limits of FIRST (1.5 mJy) and NVSS (2.5 mJy). Limited to the depth of FIRST and NVSS surveys, we can only confirm that  none of newly discovered quasars have $R\gtrsim500$.

In order to further constrain the average radio emission of these quasars, we stacked VLA FIRST images of 17 quasars locate within the footprint of the FIRST survey. The stacked image reached a rms of 0.037 mJy and is shown in Figure \ref{fig_stack}. We do not detect any significant radio emission from the stacked image, which limits the mean radio emission of these 17 quasars to be $\lesssim$0.1 mJy at a 3-$\sigma$ level. Future deeper radio imaging of these $z>6.5$ quasars are required to identify radio loud $z>6.5$ quasars and further study their radio emissions.

\section{Quasar Luminosity Function at $z\sim6.7$}  \label{sec_qlf}
\subsection{A Complete Quasar Sample}\label{qsosample}
Since most of our discoveries are from our main sample, we will calculate the QLF only using the main quasar sample. There are 15 quasars reported in this paper from the main sample, two quasars reported in Paper I and Paper II, and another 5 quasars discovered by PS1 high redshift quasar survey \citep{venemans15,mazzucchelli17} that also satisfy our selection criteria. The previously known $z\gtrsim 6.4$ quasars recovered by our main selection are listed in Table \ref{samplez7qso}. Due to the non-detection requirement in $g,r,i$ bands (Eq. 1) and $z$-dropout cut (Eq. 5) in our selection procedure, our selection is highly incomplete at $z\lesssim6.4$ (See Section \ref{selfunc} for more details). Thus we further reject three newly discovered quasars (J1535+1943, J0803+3138 and J1316+1028) and two additional quasars (P261+19 and P183+05) at $z<6.45$ from \cite{mazzucchelli17} when calculating the QLF. Our main sample missed four $z>6.4$ quasars (P011+09, P167--13, P231--20, and P006+39) discovered by the PS1 quasar surveys \citep{venemans15,tang17,mazzucchelli17}, because they do not fall into the DELS footprint. 

The final quasar sample we used for the QLF measurement includes 17 quasars with $6.45<z<7.05$ and $-27.6<M_{1450}<-25.5$. The redshifts of these quasars were measured from Ly$\alpha$ and \Mgii\ broad emission lines or from [\Cii ] emission lines in sub-millimeter \citep{decarli18} with a median redshift of $\langle z \rangle \sim6.7$. The apparent and absolute AB magnitudes of continuum at rest-frame 1450\AA\ ($m_{1450}$ and $M_{1450}$) for those recovered known quasars are derived using the same method  described in Section \ref{discoveries}. 

\subsection{Area Coverage}\label{area}
Since the PS1 covers the whole DELS footprint and we do not reject any sky area that is not covered by the NIR surveys (i.e. a small area at $\rm Decl.>60^\circ$), the total searching area is basically the footprint of DELS DR4 and DR5. We require quasar candidates to have at least one DELS $z$-band observation. But we do not limit our selections with $g$ and/or $r$ bands observations, thus the $g$- and $r$-bands observations do not affect our sky coverage estimate. The DELS DR4 covers 3,267 deg$^2$ in $z$-band\footnote{\url{http://legacysurvey.org/dr4/description/}} and the DELS DR5 covers 9972 deg$^2$ in $z$-band\footnote{\url{http://legacysurvey.org/dr5/description/}}. However, there are $\sim200$ deg$^2$ overlap regions observed, thus we need to avoid double-counting the overlap regions when estimating the area coverage. Instead of adding the DR4 and DR5 coverage, we generate a photometric catalog including both DELS DR4 and DR5 photometric data and only keep one object if it has duplicated detections in DR4 and DR5. We then use the Hierarchical Equal Area isoLatitude Pixelization \citep[HEALPix;][]{gorski05} to estimate the sky coverage of DR4+DR5 following \cite{jiang16}. The final estimated sky area by HEALPix is 13,020 deg$^2$, which is consistent with the area estimated by adding DR4 and DR5 coverage and removing overlap regions.

\subsection{Selection Function}\label{selfunc}
Following our previous works \citep{mcgreer13,mcgreer18,yang16,jiang16}, we use simulations to estimate the completeness of our selection procedure, including the color cuts and flux limits that we applied in Section \ref{zselection}. The simulation is performed under the assumption that the shape of quasar SEDs does not evolve with redshift. We generate a grid of model quasars using the simulations by \cite{yang16}, which is an updated version of the simulations by \cite{mcgreer13}. The modeled quasar spectra are designed to match the colors of $\sim$60,000 SDSS BOSS quasars in the redshift range of $2.2<z<3.5$ \citep{ross12}. Each quasar spectrum consists of a broken power-law continuum, a series of emission lines with Gaussian profiles, and Fe emission templates \citep{boroson92,vestergaard01,tsuzuki06}. The distributions of spectral features, such as the continuum spectral slope, line equivalent width (EW), and line full width at half maximum (FWHM) are matched to that of BOSS quasars. 

The simulated spectra also involve absorptions from neutral hydrogen absorption in Ly$\alpha$ forest. Finally, photometry is derived from simulated quasars and photometric errors are added for each survey by matching the observed magnitude and error relations with a large representative point source sample \citep{yang16}. The PS1 coverage depends on the sky position and the depth is not uniform. \cite{chambers16} gives the all-sky distribution of magnitude limits for 50\% and 98\% completeness on the PS1 3$\pi$ stacked data, which indicates that PS1 DR1 catalog is $\sim$50\% complete for $y_{ps1}\gtrsim21.5$ objects and is $\gtrsim$98\% complete for  $y_{ps1}<21$ objects (except for some low galactic latitude regions). 
We need to consider the sky position-dependent photometric uncertainties caused by the PS1 inhomogeneous coverage. To correct this effect in our simulations, we mapped the PS1 spatial surveying depth within our searching area and fit a 2D magnitude--coverage--error relation following the procedure explored by \cite{yang16}. Similarly, we also apply the same method for the simulated DELS photometry. We refer to \cite{mcgreer13} and \cite{yang16} for more detailed descriptions of the simulation. 

\begin{figure}
\centering
\includegraphics[width=0.5\textwidth]{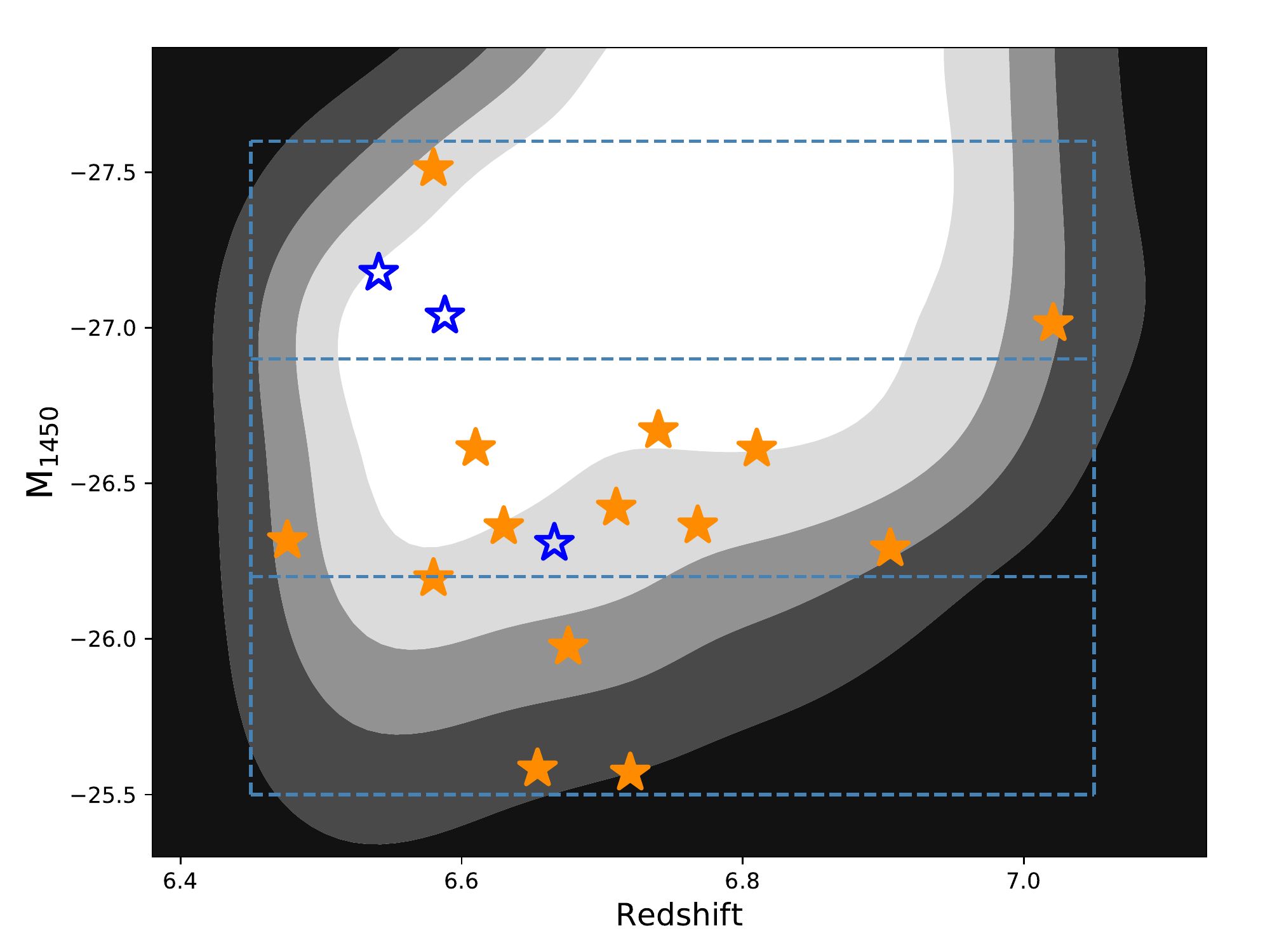}
\caption{Selection function of our $z\sim7$ quasar survey. The probability is the fraction of simulated quasars selected by our selection criteria among all simulated quasars in each ($M_{1450}$, $z$) bin. The orange solid asterisks denote newly discovered quasars reported in this work and blue open asterisks are previously known $z>6.45$ quasars. The contours are selection probabilities from 0.7 to 0.1 with an interval of 0.2. 
\label{prob}}
\end{figure}

\begin{figure}
\centering
\includegraphics[width=0.5\textwidth]{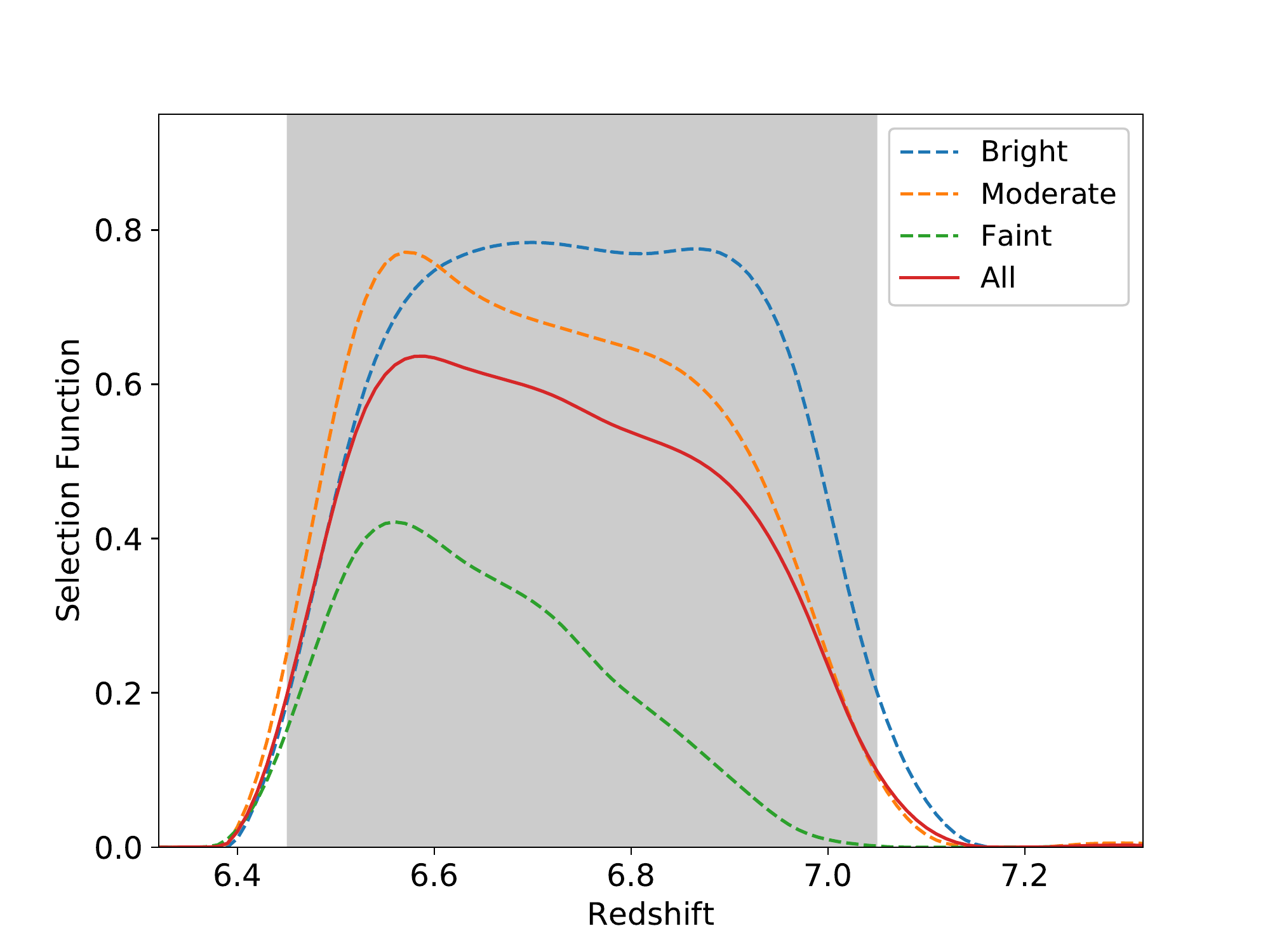}
\caption{Selection function as a function of redshift for three different luminosity bins. The blue, orange and green dashed lines denote selection function for quasars with $-27.6<M_{1450}<-26.9$, $-26.9<M_{1450}<-26.2$, and $-26.2<M_{1450}<-25.5$, respectively. The red solid line represents the selection function over the whole magnitude range. The shaded region shows the redshift bin we used for calculating QLF. \label{prob_z}}
\end{figure}

We use the simulation described above to estimate the completeness of our selection criteria. To derive a selection function, we construct a grid of simulated quasars distributed evenly in ($M_{1450}$, $z$) space with 100 quasars bin$^{-1}$ of $\Delta M_{1450}$=0.1 and $\Delta z$=0.05. Then we compute the average selection probability, $p({M_{1450},z})$, in each ($M_{1450}$, $z$) bin. The computed selection function in the ($M_{1450}$, $z$) space as well as the seventeen quasars we used for QLF measurement are shown in Figure \ref{prob}. We note that there are two faint quasars have probabilities below 30\%: J0910+1656 at $z=6.72$ and J1216+4519 at $z=6.654$. In particular, J0910+1656 has a probability of only $\sim$10\%. The reason we can select J0910+1656 is that this quasar has a strong Ly$\alpha$ emission and thus is bright in both $y_{\rm ps1}$ and $z_{\rm DELS}$, although it is faint in rest-frame 1450\AA. J1216+4519 is not that extreme but has similar situation with J0910+1656. J1216+4519 is bright in $y_{\rm ps1}$ and $z_{\rm DELS}$ but faint in $M_{1450}$. There is no any quasar found in the highly complete region at redshift between 6.6 and 6.9, which is probably because there is no such bright quasar in our searching area at this redshift range. Not surprising, this is because the number density of such luminous quasar at $z>6.6$ is very low. For example, there are only three $z>6.5$ quasars with $M_{1450}$ brighter than --27 previously known at $z>6.5$ with two of them are recovered by our selection the other one is out of our searching sky area \citep{mazzucchelli17}.

Our quasars span a magnitude range of $M_{1450}$ from --27.51 to --25.51. In order to include a statistical quasar sample in each magnitude bin, we divide our sample into 3 magnitude bins with $\Delta M_{1450}$=0.7 mag over the magnitude range --25.5$<M_{1450}<$--27.6. Figure \ref{prob_z} shows the selection function as a function of redshift in three different luminosity bins. As expected from the $z$-dropout cut, our main selection procedure has a very sharp change of completeness at $z\sim6.45$. Because we required strong detections in both PS1 $y_{ps1}$ and DELS $z$-band, our main selection limits the quasar redshift to be lower than $7.05$ and can only select very luminous quasars at $z\gtrsim 6.8$. For these reasons, we choose the redshift range from 6.45 to 7.05 (the shaded region in Figure \ref{prob_z}) when calculating the QLF. As shown in Figure \ref{prob}, the main selection procedure misses very bright quasars at $z\lesssim6.6$ due to requirements of non-detections in PS1 $i$-band.

\subsection{Spectroscopic Completeness}\label{specprob}

\begin{figure}
\centering
\includegraphics[width=0.5\textwidth]{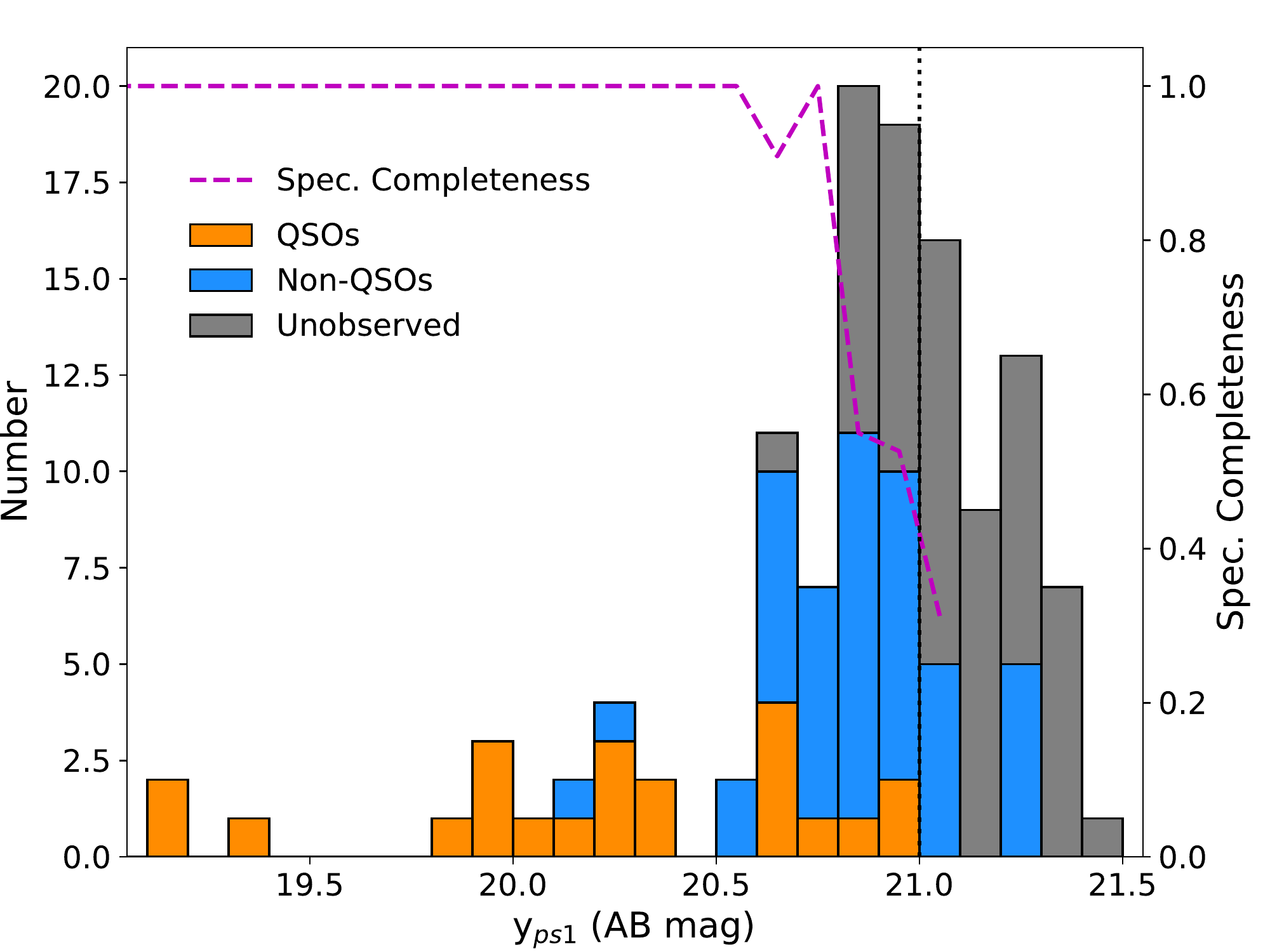}
\caption{Spectroscopic completeness of our main sample. The magenta dashed line denotes the spectroscopic completeness as a function of $y_{ps1}$-band magnitude. The histogram is divided into several components filled by different colors and represents newly identified high-redshift quasars (orange), non-quasars (blue), and unobserved candidates (gray). The black dotted line represents $y_{ps1}$=21.0 mag, which we treat as our quasar survey flux limit.
\label{spec_comp}}
\end{figure}

As mentioned in Section \ref{sec_obs}, we spectroscopically observed 65 main candidates. These 65 observed targets include quasars J1048--0109, and J0038--1527, which have been published in Paper I and Paper II. There are four candidates in our spectroscopically observed sample that can neither be rejected nor confirmed as high redshift quasars based on available spectra, and we can not count these four candidates as spectroscopically observed targets. Thus, the overall success rate of our main selection is 29.5\% (18/61). The success rate at $y_{ps1}<20.5$ is very high (10/12=83.3\%) and declines rapidly towards fainter objects because the dropout bands are not deep enough for fainter candidates. There are five more previously known quasars, J0226+0302, J1212+0505, J1724+1901, J2132+1217, and J2232+2930 that can also be treated as spectroscopically observed targets. So, 66 out of 121 main sample candidates were spectroscopically observed in total. The $y_{ps1}$ magnitude distribution of our observed and unobserved candidates is shown in Figure \ref{spec_comp}. The number of spectroscopically observed candidates is a function of $y_{ps1}$-band magnitude, which is used to correct the incompleteness by assuming the probability of an unobserved candidate to be a quasar is the same as in the observed sample at a certain magnitude. As shown in Figure \ref{spec_comp}, we do not identify any high redshift quasar at $y_{ps1}>21.0$, which means we can not correct the quasar fraction for unobserved candidates at this magnitude range. Figure \ref{spec_comp} shows that the number of main candidates drops very fast at $y_{ps1}>21.0$, which is caused by the fact that the depth of PS1 is not uniform and faint objects can only be detected at 7-$\sigma$ level in deep regions. Considering these two limitations, we only use the $y_{ps1}<21.0$ main sample to calculate the QLF.

\begin{figure}
\centering
\includegraphics[width=0.5\textwidth]{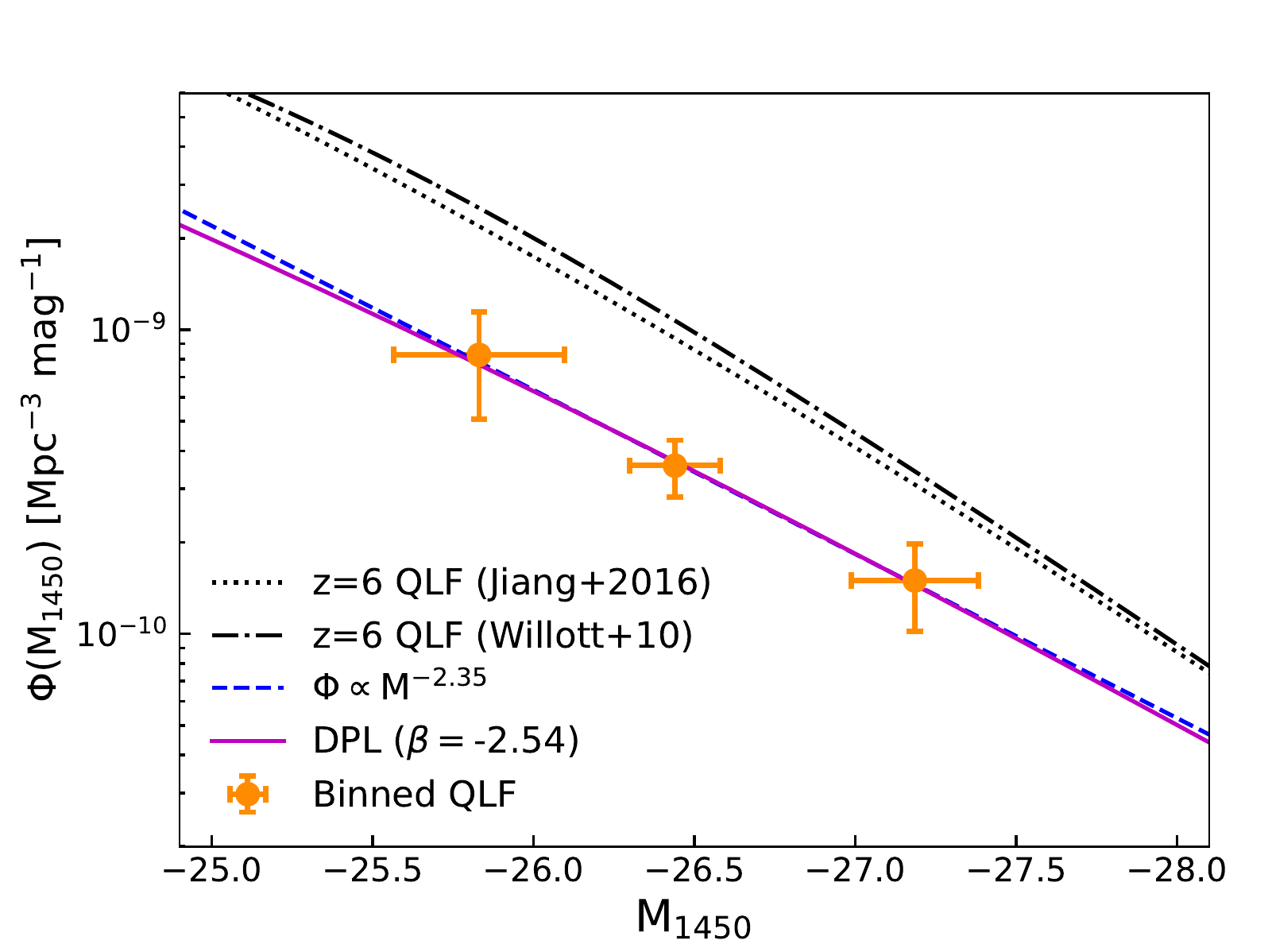}
\caption{Binned quasar luminosity function. The orange solid symbols represent our newly measured binned QLF at $z\sim6.7$. The blue dashed line is our best fit with $\beta$=$-2.35\pm0.22$. The magenta solid line denotes double power-law fit. The black dotted and dash-dotted lines represent $z\sim6$ QLF measured by \cite{jiang16} and \cite{willott10}, respectively. Clearly, the quasar number density at $z\sim6.7$ is much lower than that at $z\sim6$.
\label{qlf}}
\end{figure}

\subsection{Luminosity Function at $z\sim6.7$}\label{binqlf}
To compute the binned QLF, we divide our sample into three magnitude bins as mentioned in Section \ref{selfunc}. Due to the narrow redshift interval of our sample and the small number of high redshift quasars, we only use one redshift bin and do not take into account any redshift evolution within this redshift range. The volume densities of quasars are calculated using the standard 1/$V_{a}$ method \citep{page00}, after all incompleteness corrections have been applied for each quasar.

The binned QLF is shown in Figure \ref{qlf}. 
The QLF can be well characterized by a single power-law, $\Phi(L_{1450})\propto L_{1450}^{\beta}$, or, 
\begin{equation}\label{eq_pl}
\Phi(M_{1450})=\Phi^{\ast}10^{-0.4(\beta+1)(M_{1450}+26)},
\label{bslope}
\end{equation}
where we only consider luminosity dependence but ignore redshift evolution over our narrow redshift range. The best fits are 
$\Phi^{\ast}=(6.34\pm1.73) \times 10^{-10}$ Mpc$^{-3}$ mag$^{-1}$ and 
$\beta=-2.35\pm0.22$.

At lower redshifts, QLFs are commonly characterized using a double power-law,
\begin{equation}\label{eq_dpl}
\Phi(M,z) \\
	=\frac{\Phi^{\ast}(z)} 
   {10^{0.4(\alpha+1)(M-M^{\ast})}+10^{0.4(\beta+1)(M-M^{\ast})}},
\end{equation}   
where $\alpha$ and $\beta$ are the faint-end and the bright-end slopes, $M^{\ast}$ is the characteristic magnitude and $\Phi^{\ast}(z)=\Phi^{\ast}(z=6)\times10^{k(z-6)}$ is the normalization. Since our binned QLF only covers a narrow luminosity range, we can not fit $\alpha$ and $M^{\ast}(z)$. Currently, there is no $z>6.5$ QLF measurement at faint-end, so we fix $\alpha$ and $M^{\ast}$ to the $z\sim6$ QLF measured by \cite{jiang16}: $\alpha=-1.90$ and $M^{\ast}=-25.2$. Here we use least-square fitting rather than maximum likelihood fitting due to the lack of faint quasars and unknown faint-end slope and characteristic magnitude. Our least-square fitting gives $\Phi^{\ast}=(3.17\pm0.85) \times 10^{-9}$ Mpc$^{-3}$ mag$^{-1}$ and $\beta=-2.54\pm0.29$.

At lower redshifts ($z\lesssim5$), the QLF has a very steep bright-end slope $\beta \le -3$ and a flat faint-end slope $\alpha \sim -2.0 - -1.5$ \citep[e.g.][]{richards06, mcgreer13, yang16, Schindler18}. \cite{jiang16} measured a slightly flatter bright-end slope of $\beta=-2.8\pm0.2$ at $z\sim6$ with SDSS quasars brighter than $M_{1450}=-25.3$ \citep[also see][]{willott10}. However, the bright-end slope changes to be $\beta=-2.56\pm0.16$, if they fit a single power-law to all SDSS quasars ($-29.10 \le M_{1450} \le -24.3$). More recently, \cite{kulkarni18} measured a much steeper bright-end slope $\beta=-5.05_{-1.18}^{+0.76}$ at $z\sim6$. The large difference in the bright-end slope between \cite{kulkarni18} and previous works is mainly because \cite{kulkarni18} measured a very bright characteristic magnitude which is $M^\ast=-29.21$. In other words, the QLF measured by \cite{kulkarni18} follows a single power-law with a slope of $-2.41_{-0.08}^{+0.10}$ at $M_{1450}\gtrsim -29.0$, which is similar to the single power-law fitting at  $-29.10 \le M_{1450} \le -24.3$ by \cite{jiang16}. We measured the single power-law QLF slope to be $\beta=-2.35\pm0.22$ at $z=6.7$ using a sample of quasars with $-27.6 < M_{1450} < -25.5$. The slope changes to $\beta=-2.54\pm0.29$ if we fix the faint-end slope and characteristic magnitude to be the values derived by \cite{jiang16}, it changes to $\beta=-2.34\pm0.22$ if we fix the bright end slope and characteristic magnitude to be the values derived by \cite{kulkarni18}. Our result suggests that the QLF slope does not evolve strongly from $z\sim6$ \citep{jiang16,kulkarni18} to $z=6.7$ over a magnitude range of $-27.6 < M_{1450} < -25.5$.

\section{Discussion} \label{sec_discussion}
\subsection{Density Evolution of High-Redshift Luminous Quasars}\label{sec_density}
\begin{figure}
\centering
\includegraphics[width=0.5\textwidth]{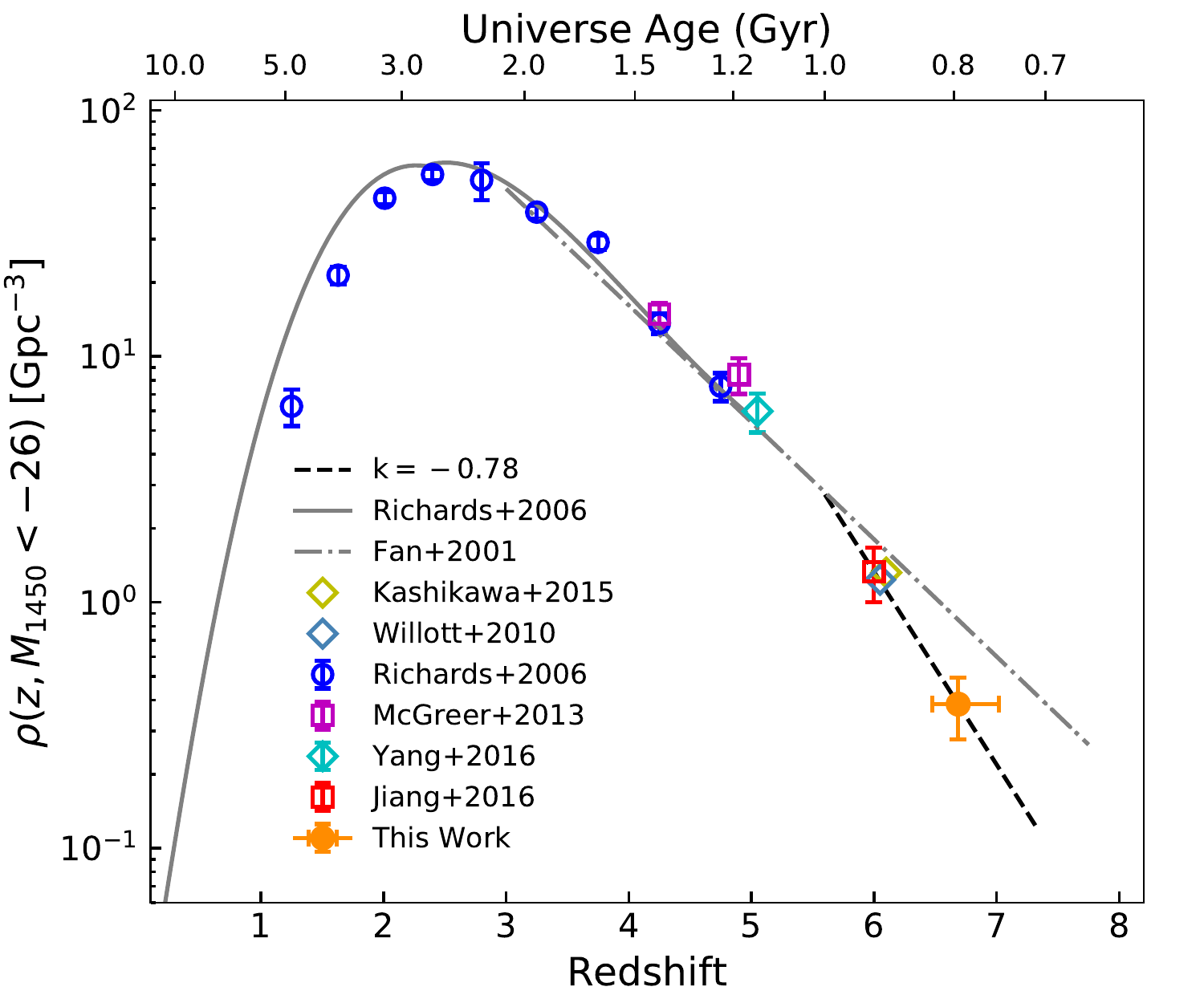}
\caption{Density evolution of luminous quasars. The grey solid line and dashed line denote evolution model from \cite{richards06} and \cite{fan01}, respectively. The black dashed line denotes density evolution model from $z\sim6$ to $z\sim6.7$ with $k=-0.78$. The grey solid line and grey dot-dashed lines are from \cite{richards06} and \cite{fan01}, respectively.
The orange solid circle denote our measurement at the highest redshift. The blue open circles are density measured from the binned SDSS quasar luminosity function \citep{richards06}. The magenta and red open squares are density measured using binned luminosity function from \cite{mcgreer13} and \cite{jiang16}, respectively. The cyan, steel blue and yellow open diamonds denote densities integrated from QLF measured by \cite{yang16}, \cite{willott10} and \cite{kashikawa15}, respectively.\label{density}}
\end{figure}

A rapid decline in the comoving number density of luminous quasars at high redshift was suggested \cite{fan01}, who fit an exponential decline to the quasar spatial density, $\rho(<M,z) \propto 10^{kz}$, and find that the density evolves from $z\sim3$ to $z\sim$6 with $k=-0.47$. This value has been frequently used in many previous works \citep[e.g.][]{willott10,kashikawa15}. An even more rapid decline in the comoving number density from $z\sim5$ to $z\sim6$ ($k\sim-0.7$) is claimed by recent studies \citep{mcgreer13,jiang16}. Here, we explore in detail the spatial density evolution of luminous quasars at higher redshifts. The spatial density of quasars brighter than a given magnitude $M$ can be calculated by integrating the QLF:
\begin{equation}
    \rho(<M,z) = \int_{-\infty}^{M}\Phi(M,z) dM ~,
 \label{eqn:rhoM}
\end{equation}

We can also estimate the density using the 1/$V_{a}$ method based on individual quasars and selection function, 
\begin{equation}
    V_{a}= \int_{\Delta z}p(M_{1450},z) \frac{dV}{dz}dz ~,
 \label{eqn:va}
\end{equation}
where $p(M_{1450},z)$ is the selection function at each magnitude and redshift bin.  The total spatial density and its uncertainty can then be estimated by:
\begin{equation}
     \rho(<M,z) = \sum_{i} \frac{1}{V_{a}^i}~,~ \sigma(\rho) = \left[\sum_{i}{\left(\frac{1}{V_{a}^{i}}\right)}^2\right]^{\frac{1}{2}},
 \label{eqn:rhova}
\end{equation}

We estimate the density of quasars brighter than $M_{1450}<$--26 at $z\sim6.7$ to be $\rm (0.39\pm0.11)~ Gpc^{-3}~mag^{-1}$, by summing over all the quasars used for QLF measurement and with $M_{1450}<$--26 using Eq. \ref{eqn:rhova}. With the same method, \cite{jiang16} measured the density at $z=6$ to be $\rm (1.33\pm0.33)~ Gpc^{-3}$ using a large sample of luminous SDSS quasars. The spatial density of quasars at $z=6.7$ is more than three times lower than that at $z=6$. In Figure \ref{density}, we show the estimated quasar spatial density at $z\sim6.7$, together with the results at $z<5$ from \cite{richards06}, $z\sim4$--5 from \cite{mcgreer13}, $z\sim6$ from \cite{jiang16} and values derived by integrating the QLFs \citep[e.g.][]{yang16,willott10,kashikawa15} using Eq. \ref{eqn:rhoM}. 

We derive the exponential density evolution parameter to be $k=-0.78\pm0.18$ from $z\sim6$ to $z\sim6.7$ by fitting our newly estimated density at $z\sim6.7$ and the density at $z\sim6$ from \cite{jiang16}. Although \cite{mcgreer13} and \cite{jiang16} have noticed that the decline of quasar number density is accelerating at $z\sim6$ (also see Figure \ref{density}), our findings suggests that the decline is even more rapidly towards higher redshift. Such steep decline rate indicates that the spatial density of luminous quasars drops by a factor of $\sim6$ per unit redshift towards earlier cosmic epoch, two time faster than that at $z\sim3-5$. If such decline extends to higher redshift, we expect to see only one such luminous quasar over the whole visible sky at $z\sim9$. This means that we are finally witnessing the first quasars in the EoR and we will be badly limited by the small number of such quasars when studying the reionization history and SMBH growth history.

Quasar evolution at $z>6$ is limited by the number of {\it e}-folding times available for BH accretion. The rapid decline of luminous quasar spatial density within such short cosmic time (i.e., $\sim$121 Myrs, or three {\it e}-folding times) requires that SMBHs must grow rapidly from $z\sim6.7$ to $z\sim6$ or they are less radiatively efficient at $z\sim6.7$. For J0706+2921, the brightest quasar in our sample, it takes 20 {\it e}-folding times, or the age of the universe at $z\sim6.6$, to grow from a 10 $\rm M_\odot$ stellar black hole, assuming the radiation efficiency $\epsilon$=0.1. The existence of these luminous quasars helps determine whether standard models of radiatively efficient accretion from stellar seeds are still allowed, or alternative models of BH seed formation and BH accretion (super-Eddington or radiatively inefficient) are required \citep[e.g.][]{volonteri06}.

In addition, the determination of luminous quasar spatial density evolution at high redshift has important consequences in understanding early BH growth and BH-galaxy co-evolution \citep[e.g.][]{wyithe03,hopkins05,shankar10}. 
Combining the dark matter halo mass and duty circle inferred from quasar clustering measurements \citep{shen07}, \cite{shankar10} predict the QLF at $z > 3$ and claim that the rapid drop in the abundance of massive and rare host halos at $z \gtrsim 7$ implies a proportionally rapid decline in the number density of luminous quasars, much stronger than simple extrapolations of the $z = 3-6$ luminosity function. Our measurement is consistent with their prediction, with a even stronger declining rate of quasar number density, which requires that these luminous $z>6.5$ quasars reside in even more massive ($ M_{\rm halo}\gtrsim10^{13}M_\odot$), less numerous dark matter halos than that of luminous $z\sim6$ quasars.  However, there is currently no any conclusive direct observations on whether these high redshift quasars reside in the most massive dark matter halos. Future deep wide field imaging and spectroscopy are needed to show whether luminous quasars reside in the most biased environment.

\subsection{Quasar Contribution to Reionization}\label{reionization}

\begin{figure}
\centering
\includegraphics[width=0.5\textwidth]{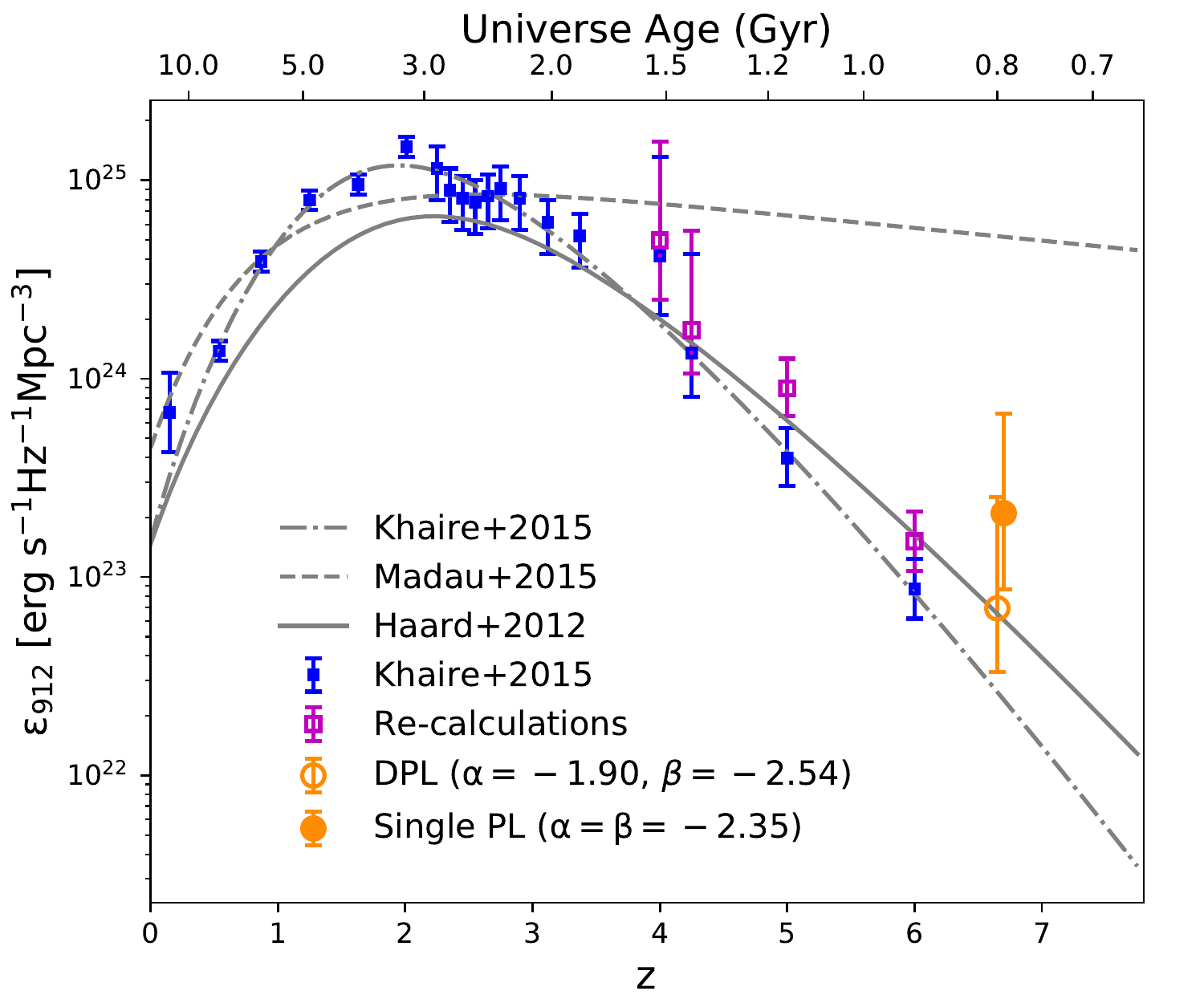}
\caption{The quasar comoving emissivity at 912 \AA\ ($\epsilon _{912}$) versus redshift ($z$). The small blue  squares are taken from the compilations of \cite{khaire15} by integrating QLFs down to 0.01 $L_{\ast}$. The magenta open squares are our recalculations of their $z>4$ points by integrating QLFs down to --18 mag and assuming a broken power-law quasar SED mentioned in the main text (i.e. same values used for $z\sim6.7$ quasar emissivity measurement). The solid and open circles are our calculations at $z\sim6.7$ by using single power-law QLF and double power-law QLF, respectively. Note that the open circles are shifted by 0.05 on x-axis for clarity.
The solid, dash-dotted and dashed lines are models from \cite{haardt12}, \cite{khaire15} and \cite{madau15}.\label{emissivity}}
\end{figure}

Here, we estimate the quasar contribution to the ionizing photons at $z\sim6.7$ based on our newly derived QLF. We first calculate the quasar emissivities by assuming a broken power-law quasar SED with an index of $\alpha_\nu=-1.7$ at ultraviolet wavelengths, a break at 912\AA, and an index of $\alpha_\nu=-0.6$ at longer wavelengths \citep{lusso15} and assuming that the escape fraction is 100\%. As the characteristic magnitude of QLF at high redshift is highly uncertain, we integrate our QLF down to --18 mag following previous works \citep{kashikawa15,yang16,jiang16}, rather than integrate QLF down to $\sim$0.01 M$^{\ast}$ \citep{madau99, khaire15}. 

As we have measured $\Phi=(6.34\pm1.73) \times 10^{-10}$ Mpc$^{-3}$ mag$^{-1}$ at  $M_{1450}=-26.0$ and the QLF slope to be $\beta=-2.35\pm0.22$ at $-27.6<M_{1450}<-25.5$, the ionization photon production rate from quasars at $z=6.7$ are mainly determined by the QLF faint-end slope and the characteristic magnitude. First we assume that the QLF at $z\sim6.7$ is a single power-law as described by Eq. (\ref{eq_pl}). The QLF faint-end slope at $z\sim6$ was suggested to be $-2.0\lesssim \alpha \lesssim -1.5$ \citep[e.g.][]{willott10,jiang16,kashikawa15}. If the QLF faint-end slope does not evolve from $z\sim6$ to $z\sim6.7$, the single power-law with $\beta=-2.35$ used here will give a maximum emissivity measurement because our single power-law QLF will overestimate the number of quasars at the faint-end. In this case, the double power-law QLF described by Eq. (\ref{eq_dpl}) with $\alpha$ and characteristic magnitude fixed to the values determined by \cite{jiang16} are better suited to estimate quasar comoving emissivity at $z\sim6.7$. If the break magnitude evolves following the luminosity evolution and density evolution (LEDE) model \citep{mcgreer13,yang16}, the break magnitude at $z\sim6.7$ will be brighter than --27, which is also consistent with the recent QLF measurement at $z=6$ \citep{kulkarni18}, our best-fit slope is actually the QLF faint-end slope. In this case, the single power-law we used here gives the true emissivity measurement. But note that, the total quasar comoving emissivity is sensitive to the lower bound of the integral, especially for the single power-law case. Thus, the numbers estimated based on single power-law need to be cautionary use. 

The quasar comoving emissivity at 1 Ryd is estimated to be $\epsilon _{912}=(2.10_{-1.23}^{+4.54})\times10^{23}$ $\rm ergs~s^{-1}Hz^{-1}Mpc^{-3}$ and $\epsilon _{912}=(6.95_{-3.64}^{+18.4})\times10^{22}$ $\rm ergs~s^{-1}Hz^{-1}Mpc^{-3}$, by using the best fit single power-law QLF ($\beta=-2.35$) and double power-law QLF ($\alpha=-1.90$, $\beta=-2.54$, and $M^{\ast}=-25.2$), respectively (Figure \ref{emissivity}). To compare with previous works, we also over plot quasar emissivity models \citep{haardt12,khaire15,madau15} as well as measurements at lower redshifts in Figure \ref{emissivity}.  The integrated emissivity of ionizing photons from quasars at $z\sim6.7$ is then estimated to be $\dot{\cal N}_{\rm ion}=(1.86_{-1.09}^{+4.03})\times10^{49}$ $\rm photons\, Mpc^{-3}\, s^{-1}$ and $\dot{\cal N}_{\rm ion}=(6.17_{-3.23}^{+16.4})\times10^{48}$ $\rm photons\, Mpc^{-3}\, s^{-1}$ for the best-fit single power-law QLF and double power-law QLF, respectively.

\begin{figure}
\centering
\includegraphics[width=0.5\textwidth]{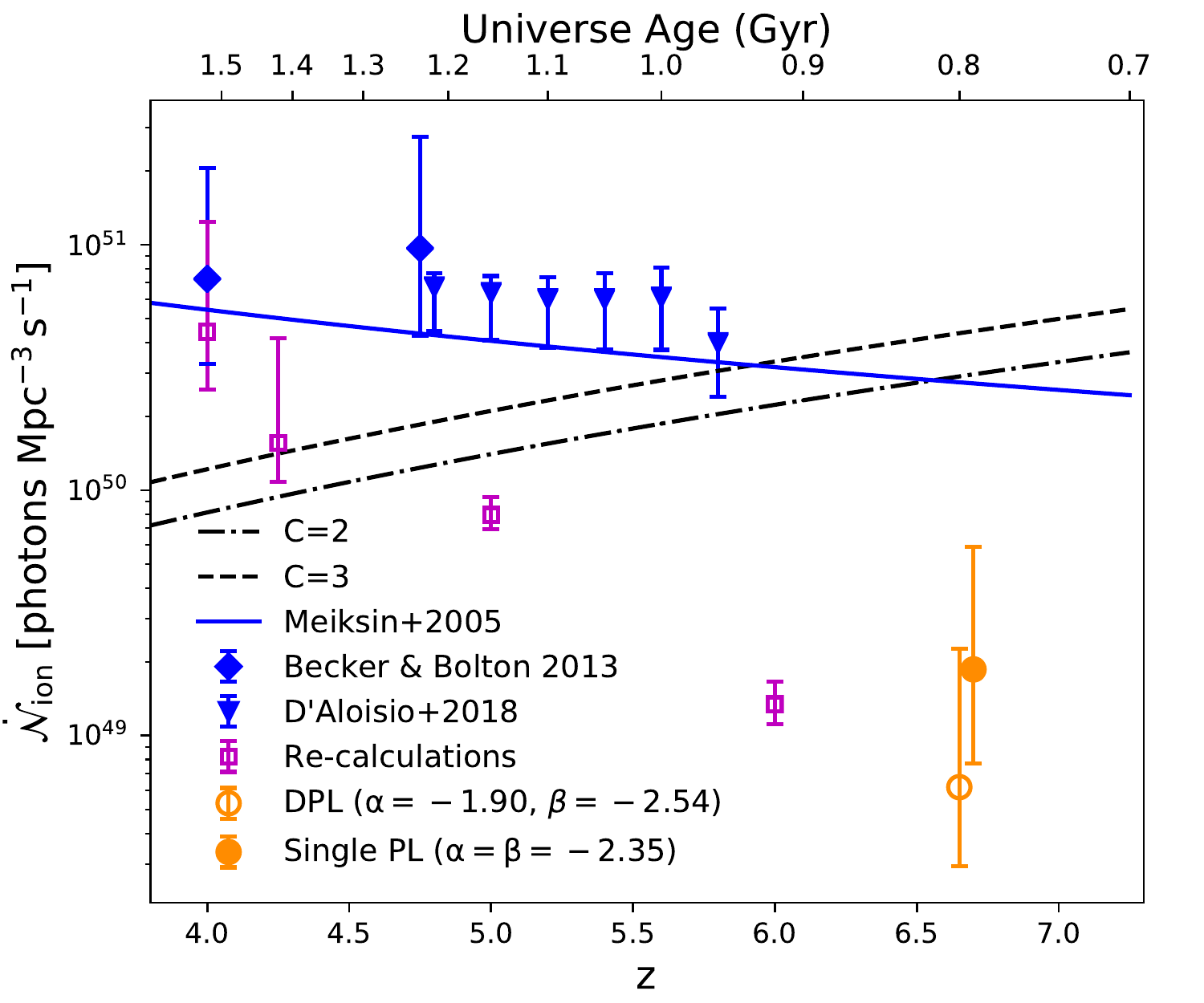}
\caption{The evolution of the comoving production rate of ionizing photons. The solid and open circles are production rate of ionizing photons at $z\sim6.7$ based on our single power-law QLF and double power-law QLF, respectively. Note that the open circles was shifted by 0.05 on x-axis for clarity. The magenta open squares are quasar ionizing photon production rate at lower redshift calculated based on our recalculations of quasar comoving emissivity from \cite{khaire15}. The blue solid diamonds and triangles denote the required photon production rate inferred from measurements of the mean transmitted Ly$\alpha$ flux by \cite{becker13} and \cite{daloisio18}, respectively. The blue solid line represent the required rate inferred from measurements of the mean transmitted Ly$\alpha$ flux from \cite{meiksin05}. The dash-dotted and dashed lines are the required photon rate density to balance hydrogen recombination by assuming  $C=2$ and $C=3$ \cite{madau99}, respectively.\label{fig_nion}}
\end{figure}

The total required photon rate density to balance hydrogen recombination was estimated by \cite{madau99}, i.e., 
$\dot{\cal N}_{\rm ion}(z) = 10^{51.16}\,
     \left(\frac{C}{3}\right) \times \left(\frac{1+z}{8}\right)^3 \left(\frac{\Omega _b h_{70}^2}{0.08}\right)^2 
   {\rm photons\, Mpc^{-3}\, s^{-1}},$
where we adopt $\Omega_b=0.047$ and $h_{70}=1.0$ in the following calculations.
The clumping factor $C$ is suggested to be $\sim$2--3 at $z\sim6-7$ by simulations with radiative transfer to capture self-shielding for ionizing backgrounds in the range of those allowed \citep{mcquinn11,kaurov15}. For $C = 2$, the total required photon rate density at $z=6.7$ is $2.97\times10^{50}~{\rm photons\, Mpc^{-3}\, s^{-1}}$ and quasars provide $\sim$ 6.3\% and $\sim$ 2.1\% of the required photons for single power-law QLF and double power-law QLF, respectively. While for $C = 3$, the required photon rate density changes to $4.45\times10^{50}~{\rm photons\, Mpc^{-3}\, s^{-1}}$ and quasars only provide $\sim$ 4.2\% and $\sim$ 1.4\% required photons for single power-law QLF and double power-law QLF, respectively. 

The required photoionization rate can also be inferred directly from measurements of the mean transmitted Ly$\alpha$ flux from quasar spectra \citep[e.g.][]{meiksin05,becker13,davies18a,daloisio18}. 
It is related to the total (comoving) emissivity of the sources, $\epsilon _L^S$, and can be described by 
$\dot{\cal N}_{\rm ion}^S= \int_{\nu_L}^{\infty} \frac{\epsilon _L^S}{h_P} (\frac{\nu}{\nu_L})^{-\alpha_s} \frac{d\nu}{\nu} \approx \frac{\epsilon_L^S}{h_P \alpha_s}$
$=A_S \frac{3+\alpha_{MG}}{3\alpha_S} (1+z)^\gamma h $ $\rm photons\, Mpc^{-3}\, s^{-1}$, where $h_P$ is the Planck constant, $A_S=1.3\times10^{52}$, $\alpha_{MG}=1$, $\gamma=-1.6$ \citep{meiksin05}. 
By assuming $\alpha_s=1.7$ \citep{lusso15}, we measured that quasars only provide $\sim$ 6.8\% and $\sim$ 2.3\% of the required photons for single power-law QLF and double power-law QLF, respectively. The evolution of the comoving production rate of ionizing photons is shown in Figure \ref{fig_nion}. 
Although we can not constrain the faint-end slope and characteristic magnitude of the QLF, the faint-end slope can not be steeper than the bright-end slope. The ionizing photon production rate estimated by using the single power-law QLF gives either a maximum estimate (if $M^\ast>-25.5$, and --2.35 is the bright-end slope) or the true value (if $M^\ast<-27.6$, and --2.35 is the faint-end slope). Based on these results, it is highly unlikely that high redshift quasars make a significant contribution to hydrogen reionization. As we mentioned before, the total quasar comoving emissivity is very sensitive to the lower bound of the integral for the single power law case. If there is a significant population of extremely faint AGN (i.e. $M_{1450}\gtrsim-18$), they would provide a significant contributions to the  hydrogen reionization.

\section{Summary} \label{sec_summary}
In this paper, we presented the discovery of 16 quasars at $z\gtrsim6.4$, and 5 quasars at $z\sim6$ using  modified color selection procedures from Paper I by adding PS1 photometry. Our newly discovered $z\gtrsim6.4$ quasars span an absolute magnitude range from $\sim$--25.5 to $\sim$--27.5. We have more than doubled the number of known luminous quasars at $z>6.5$ and constructed the first large uniformly selected quasar sample at such redshift. The statistically uniform $z\gtrsim6.5$ quasar sample constructed in this paper allows us for the first time to estimate the BAL quasar fraction at the EoR, which is $\gtrsim$22\%, slightly higher than that at lower redshift. However, future NIR spectroscopy and larger quasar sample are needed in order to finalize whether BAL quasar fraction evolves with redshift. 

After determining the completeness of the selection function estimated using simulated quasars and correcting for spectroscopic incompleteness, we calculated the QLF using a uniform sample of 17 quasars at $6.45< z < 7.05$ covering a sky area of $\sim 13000$ deg$^2$ to a flux limit of $z_{PS1} = 21$. We measured the slope of the QLF to be $\beta = -2.35\pm0.22$ within the magnitude range of $-27.6<M_{1450}<-25.5$ by fitting a single power-law. If we fit the binned QLF with a double power law and fix the faint-end slope ($\alpha=-1.9$) and characteristic magnitude ($M^\ast=-25.2$), the bright-end slope is measured to be $\beta=-2.54\pm0.29$. If we fix the bright-end slope ($\beta=-5.05$) and characteristic magnitude ($M^\ast=-29.21$), the faint-end slope is measured to be $\beta=-2.34\pm0.22$. These measurements indicate that the QLF slope does not evolve significantly from $z=6$ to $z=6.7$ within $-27.6<M_{1450}<-25.5$.

We measured the quasar spatial density at $z\sim6.7$ to be $ \rho(<-26.0)$=$\rm (0.39\pm0.11)~ Gpc^{-3}~mag^{-1}$ using the  1/$V_a$ method. By fitting the spatial density at $z=6$ and $z=6.7$, we find a density evolution parameter of $k=-0.78\pm0.18$ from $z\sim6$ to $z\sim6.7$, which corresponds to the quasar number density declining by a factor of more than three from $z\sim6$ to $z\sim6.7$, at a rate significantly faster than the average decline rate between $z\sim 3$ and 6. The cosmic time between $z\sim6$ and $z\sim6.7$ is only 121 Myrs, or less than three $e$-folding times. The rapid decline of quasar density within such short time requires that SMBHs must grow rapidly from $z\sim6.7$ to $z\sim6$ or they are less radiatively efficient at $z\sim6.7$.

We estimated $z\sim6.7$ quasar comoving emissivity at 1 Ryd to be $\epsilon _{912}=(2.10_{-1.23}^{+4.54})\times10^{23}$ $\rm ergs~s^{-1}Hz^{-1}Mpc^{-3}$ and $\epsilon _{912}=(6.95_{-3.64}^{+18.4})\times10^{22}$ $\rm ergs~s^{-1}Hz^{-1}Mpc^{-3}$, using the best fit single power-law QLF ($\beta=-2.35$) and double power-law QLF ($\alpha=-1.90$, $\beta=-2.54$), respectively. The integrated emissivity of ionizing photons from quasars at $z=6.7$ is estimated to be $\dot{\cal N}_{\rm ion}=(1.86_{-1.09}^{+4.03})\times10^{49}$ $\rm photons~s^{-1}Mpc^{-3}$ and $\dot{\cal N}_{\rm ion}=(6.17_{-3.23}^{+16.4})\times10^{48}$ $\rm photons~s^{-1}Mpc^{-3}$ for the best fit single power-law QLF and double power-law QLF, respectively. By comparing with the required ionization photon production rate estimated from mean transmitted Ly$\alpha$ flux, our measurements suggest that high redshift quasars have very low possibility to be the dominant contributor of hydrogen reionization (i.e. only contribute $<$7\% required ionizing photons).

We are collecting high quality optical and NIR spectra with large ground-based telescopes. In the forthcoming publications we will give measurements of Gunn-Peterson optical depths based on both Ly$\alpha$ and Ly$\beta$ forests, search damping wing signatures to constrain the evolution of neutral fraction at the EoR, provide the BH mass measurements and explore the SMBH growth history with the high quality spectra that we are collecting. 
We are also observing X-ray emissions from these quasars with {\it Chandra}, which will give insights the accretion status of of the earliest SMBHs. This dedicated dataset will be an ideal dataset for investigating the cosmic reionization history and SMBH formation mechanisms. We are also surveying the [\Cii] emission line from the quasar host galaxies with {\it ALMA} and {\it NOEMA}. These observations will provide us excellent opportunities to study the SMBH and host galaxy co-evolution in the EoR. Moreover, this unique quasar sample will provide ideal targets for {\it JWST} to investigate the environments and host galaxies of these SMBHs.

\acknowledgments
F. Wang, J. Yang, X.-B. Wu and L. Jiang thank the supports by the National Key R\&D Program of China (2016YFA0400703) and the National Science Foundation of China (11533001, 11721303).
J. Yang, X. Fan, M. Yue, J.-T. Schindler and I. D. McGreer acknowledge support from the US NSF Grant AST-1515115 and NASA ADAP Grant NNX17AF28G. 
J.R. Findlay, B.W. Lyke, A.D. Myers and E. Haze Nunez acknowledge support from the National Science Foundation through REU grant AST-1560461.

We acknowledge the use of data obtained at the Gemini Observatory (NOAO program ID: GN-2018A-C-1), which is operated by the Association of Universities for Research in Astronomy (AURA) under a cooperative agreement with the NSF on behalf of the Gemini partnership: the National Science Foundation (United States), the National Research Council (Canada), CONICYT (Chile), the Australian Research Council (Australia), Ministério da Ci\^encia e Tecnologia (Brazil) and Ministerio de Ciencia, Tecnolog\'ia e Innovaci\'on Productiva (Argentina).
We acknowledge the use of Hale telescope, Keck I telescope, LBT telescopes, Magellan telescopes, MMT 6.5 m telescope and UKIRT telescope. Observations obtained with the Hale Telescope at Palomar Observatory were obtained as part of an agreement between the National Astronomical Observatories, Chinese Academy of Sciences, and the California Institute of Technology. Observations reported here were obtained at the MMT Observatory, a joint facility of the University of Arizona and the Smithsonian Institution.

We acknowledge the use of BASS, DECaLS, MzLS, PanStarrs, UHS, ULAS and VHS photometric data. This publication makes use of data products from the Wide-field Infrared Survey Explorer, which is a joint project of the University of California, Los Angeles, and the Jet Propulsion Laboratory/California Institute of Technology, and NEOWISE, which is a project of the Jet Propulsion Laboratory/California Institute of Technology. WISE and NEOWISE are funded by the National Aeronautics and Space Administration.

This research uses data obtained through the Telescope Access Program (TAP), which has been funded by the National Astronomical Observatories, Chinese Academy of Sciences, and the Special Fund for Astronomy from the Ministry of Finance in China.

\facilities{Gemini (GMOS), Hale (DBSP), KECK (DEIMOS), LBT (MODS), Magellan (FIRE, LDSS3-C), MMT (BinoSpec, MMIRS, Red Channel Spectrograph), UKIRT (WFCam)}

\appendix

\section{Discovery of Five New Quasars at $z\sim6$}\label{sec_app}
We keep searching $z\sim6$ quasars using traditional $i$-dropout method by combing DELS and infrared surveys mentioned above. The detailed $i$-dropouts selection was described in \cite{wang17} and will not be repeated here. The only difference is that we used unWISE \citep{lang14, meisner17} photometry instead of ALLWISE photometry when selecting $z\sim6$ quasar candidates.

For $z\sim6$ quasar candidates, we first obtained deep $i$-band imaging with Wide-field Camera mounted on the Wyoming Infrared Observatory \citep{findlay16}. After PS1 DR1 was released, we instead used PS1 $i$-band as the dropout band. Following \cite{wang17}, we took spectroscopic observations for those candidates satisfy $i-z>2.0$ or $SN(i)<3$. We observed seven $z\sim6$ quasar candidates that passed our photometric selection with DBSP and Red Channel. The data obtained from Red Channel and DBSP were reduced using standard IRAF routines.

\begin{figure*}
\centering
\includegraphics[width=0.48\textwidth]{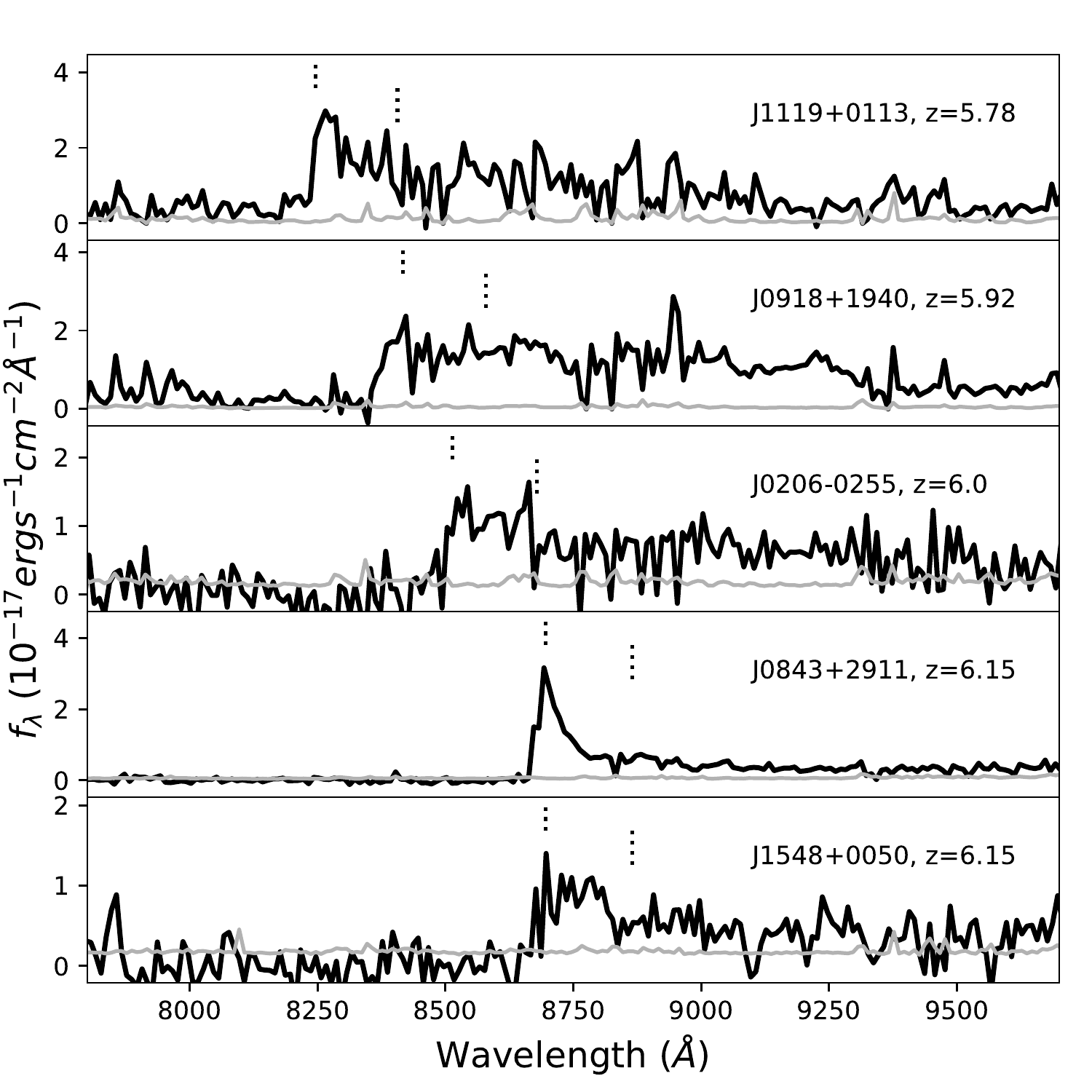}
\caption{Spectra of five newly discovered $z\sim6$ quasars. The dotted lines mark the expected positions of  $\rm Ly\alpha$ and N\,{\sc v} emission lines from left to right.
The grey lines are denote flux error vector. All spectra are binned into 10 \AA\ in wavelength space using flux error weighted mean. \label{z6spec}}
\end{figure*}

From those observed $i$-dropouts, we identified five new $z\sim6$ quasars. The observational information of these five quasars are listed in Table \ref{z6qsoobs}. Figure \ref{z6spec} shows discovery spectra of these five quasars. The redshifts of these quasars are also measured using ASERA software, with a range of $5.78\le z \le6.15$. However, due to poorer quality of these discovery spectra compared with that of $z>6.5$ quasars, the redshift uncertainties are slightly larger. As these $z\sim6$ quasars are quite faint in NIR J-band and the 1450 \AA\ more close to PS1 $y_{ps1}$-band, we scaled the composite spectra to PS1 $y_{ps1}$-band photometry to measure magnitudes at 1450 \AA . These quasars have similar luminosities with that of two quasars reported in \cite{wang17} and are fainter than majority quasars found by SDSS and PS1 quasar surveys \citep[e.g.][]{fan01,banados16}, but brighter than SHELLQs quasars \citep[e.g.][]{matsuoka16}.
Table \ref{newz6qso} present the redshift and photometric information of these five newly discovered $z\sim6$ quasars.

\begin{deluxetable*}{lccccccccccccrl}
\tabletypesize{\scriptsize}
\tablecaption{Observational information of 5 new $z\sim6$ quasars reported in this paper. \label{z6qsoobs}}
\tablewidth{0pt}
\tablehead{
\colhead{Name} & \colhead{Telescope} & \colhead{Instrument} & \colhead{Exposure (sec)} & \colhead{OBS-DATE (UT)} }
\startdata
DELS J020611.20$-$025537.8 & Hale 200 inch & DBSP & 3600 & 20160912 \\
DELS J084303.76$+$291113.4 & Hale 200 inch & DBSP & 1200 & 20170421 \\
DELS J091828.65$+$194045.0 & MMT & Red Channel & 2400   & 20170331 \\
DELS J111921.65$+$011308.6 & MMT & Red Channel & 2400   & 20170331 \\
DELS J154825.40$+$005015.5 & Hale 200 inch & DBSP & 5400 & 20170416, 20170421 \\
\enddata
\end{deluxetable*}

\begin{deluxetable*}{lccccccccccccrl}
\tabletypesize{\scriptsize}
\tablecaption{Properties of five newly discovered $z\sim6$ quasars. \label{newz6qso}}
\tablewidth{0pt}
\tablehead{
\colhead{Name} & \colhead{Redshift} & \colhead{$m_{1450}$} & \colhead{$M_{1450}$} &\colhead{$i_{\rm AB}$}&
\colhead{$z_{\rm DELS, AB}$} & \colhead{$y_{\rm ps1, AB}$} & \colhead{$J_{\rm VEGA}$} & \colhead{$W1_{\rm VEGA}$\tablenotemark{d}} & \colhead{NIR Survey}
}
\startdata
DELS J154825.40$+$005015.5 & 6.15$\pm$0.05 & 21.11$\pm$ 0.18 & --25.62$\pm$ 0.18 &24.58$\pm$0.67\tablenotemark{b} &20.82$\pm$ 0.02 & 21.12$\pm$ 0.18 & 19.79$\pm$ 0.25 & 17.32$\pm$ 0.10 & LAS\\
DELS J084303.76$+$291113.4 & 6.15$\pm$0.03 & 20.94$\pm$ 0.15 & --25.79$\pm$ 0.15 & 25.09$\pm$2.75\tablenotemark{c} &20.62$\pm$ 0.04 & 20.95$\pm$ 0.15 & 20.12$\pm$ 0.22 & 17.79$\pm$ 0.18 & LAS\\
DELS J020611.20$-$025537.8\tablenotemark{a} & 6.00$\pm$0.05 & 20.87$\pm$ 0.17 & --25.82$\pm$ 0.17 & 24.08$\pm$1.21\tablenotemark{c} &21.27$\pm$ 0.05 & 20.86$\pm$ 0.17 & 20.34$\pm$ 0.29 & 18.67$\pm$ 0.32 & VHS\\
DELS J091828.65$+$194045.0 & 5.92$\pm$0.05 & 20.76$\pm$ 0.13 & --25.91$\pm$ 0.13 & 23.32$\pm$0.26\tablenotemark{c} &20.86$\pm$ 0.07 & 20.73$\pm$ 0.13 & 19.48$\pm$ 0.21 & 16.96$\pm$ 0.07 & UHS\\
DELS J111921.65$+$011308.6 & 5.78$\pm$0.05 & 21.06$\pm$ 0.14 & --25.57$\pm$ 0.14 & 22.75$\pm$0.11\tablenotemark{b} &20.75$\pm$ 0.02 & 20.99$\pm$ 0.14 & 19.58$\pm$ 0.23 & 16.88$\pm$ 0.07 & LAS\\
\enddata
 \tablenotetext{a}{This quasar was discovered by \cite{matsuoka17} independently.}
 \tablenotetext{b}{The $i$-band photometry come from PS1 DR1 photometric catalog. }
 \tablenotetext{c}{The $i$-band photometry were obtained with Wide-field Camera mounted on the Wyoming Infrared Observatory \citep{findlay16}. }
 \tablenotetext{d}{The WISE W1 magnitudes come from unWISE photometric catalog.}

\end{deluxetable*}

\begin{thebibliography}{}
\bibitem[Allen et al.(2011)]{allen11} Allen, J.~T., Hewett, P.~C., Maddox, N., Richards, G.~T., \& Belokurov, V.\ 2011, \mnras, 410, 860 
\bibitem[Arnaboldi et al.(2007)]{arnaboldi07} Arnaboldi, M., Neeser, M.~J., Parker, L.~C., et al.\ 2007, The Messenger, 127, 28 
\bibitem[Ba{\~n}ados et al.(2016)]{banados16} Ba{\~n}ados, E., Venemans, B.~P., Decarli, R., et al.\ 2016, \apjs, 227, 11 
\bibitem[Ba{\~n}ados et al.(2018)]{banados18} Ba{\~n}ados, E., Venemans, B.~P., Mazzucchelli, C., et al.\ 2018, \nat, 553, 473 
\bibitem[Ba{\~n}ados et al.(2015)]{banados15} Ba{\~n}ados, E., Venemans, B.~P., Morganson, E., et al.\ 2015, \apj, 804, 118 
\bibitem[Becker \& Bolton(2013)]{becker13} Becker, G.~D., \& Bolton, J.~S.\ 2013, \mnras, 436, 1023 
\bibitem[Becker et al.(1995)]{becker95} Becker, R.~H., White, R.~L., \& Helfand, D.~J.\ 1995, \apj, 450, 559 
\bibitem[Beckwith et al.(2006)]{beckwith06} Beckwith, S.~V.~W., Stiavelli, M., Koekemoer, A.~M., et al.\ 2006, \aj, 132, 1729 
\bibitem[Best et al.(2015)]{best15} Best, W.~M.~J., Liu, M.~C., Magnier, E.~A., et al.\ 2015, \apj, 814, 118 
\bibitem[Bolton et al.(2011)]{bolton11} Bolton, J.~S., Haehnelt, M.~G., Warren, S.~J., et al.\ 2011, \mnras, 416, L70 
\bibitem[Boroson \& Green(1992)]{boroson92} Boroson, T.~A., \& Green, R.~F.\ 1992, \apjs, 80, 109 
\bibitem[Bosman \& Becker(2015)]{bosman15} Bosman, S.~E.~I., \& Becker, G.~D.\ 2015, \mnras, 452, 1105 
\bibitem[Byard \& O'Brien(2000)]{byard00} Byard, P.~L., \& O'Brien, T.~P.\ 2000, \procspie, 4008, 934 
\bibitem[Carilli \& Walter(2013)]{carilli13} Carilli, C.~L., \& Walter, F.\ 2013, \araa, 51, 105 
\bibitem[Carilli et al.(2007)]{carilli07} Carilli, C.~L., Wang, R., van Hoven, M.~B., et al.\ 2007, \aj, 133, 2841 
\bibitem[Casali et al.(2007)]{casali07} Casali, M., Adamson, A., Alves de Oliveira, C., et al.\ 2007, \aap, 467, 777 
\bibitem[Chambers et al.(2016)]{chambers16} Chambers, K.~C., Magnier, E.~A., Metcalfe, N., et al.\ 2016, arXiv:1612.05560 
\bibitem[Condon et al.(1998)]{condon98} Condon, J.~J., Cotton, W.~D., Greisen, E.~W., et al.\ 1998, \aj, 115, 1693 
\bibitem[D'Aloisio et al.(2018)]{daloisio18} D'Aloisio, A., McQuinn, M., Davies, F.~B., \& Furlanetto, S.~R.\ 2018, \mnras, 473, 560 
\bibitem[Davies et al.(2018b)]{davies18} Davies, F.~B., Hennawi, J.~F., Ba{\~n}ados, E., et al.\ 2018, arXiv:1802.06066 
\bibitem[Davies et al.(2018a)]{davies18a} Davies, F.~B., Hennawi, J.~F., Eilers, A.-C., \& Luki{\'c}, Z.\ 2018, \apj, 855, 106 
\bibitem[Decarli et al.(2017)]{decarli17} Decarli, R., Walter, F., Venemans, B.~P., et al.\ 2017, \nat, 545, 457 
\bibitem[Decarli et al.(2018)]{decarli18} Decarli, R., Walter, F., Venemans, B.~P., et al.\ 2018, \apj, 854, 97 
\bibitem[Dey et al.(2018)]{dey18} Dey, A., Schlegel, D.~J., Lang, D., et al.\ 2018, arXiv:1804.08657 
\bibitem[Dye et al.(2018)]{dye18} Dye, S., Lawrence, A., Read, M.~A., et al.\ 2018, \mnras, 473, 5113 
\bibitem[Faber et al.(2003)]{faber03} Faber, S.~M., Phillips, A.~C., Kibrick, R.~I., et al.\ 2003, \procspie, 4841, 1657 
\bibitem[Fabricant et al.(1998)]{fabricant98} Fabricant, D.~G., Fata, R.~G., \& Epps, H.~W.\ 1998, \procspie, 3355, 232 
\bibitem[Fan et al.(2006)]{fan06} Fan, X., Carilli, C.~L., \& Keating, B.\ 2006, \araa, 44, 415 
\bibitem[Fan et al.(2001)]{fan01} Fan, X., Narayanan, V.~K., Lupton, R.~H., et al.\ 2001, \aj, 122, 2833 
\bibitem[Fan et al.(2018)]{fan18} Fan, X., Wang, F., Yang, J.\ 2018, \apjl, submitted
\bibitem[Findlay et al.(2016)]{findlay16} Findlay, J.~R., Kobulnicky, H.~A., Weger, J.~S., et al.\ 2016, \pasp, 128, 115003 
\bibitem[Gibson et al.(2009)]{gibson09} Gibson, R.~R., Jiang, L., Brandt, W.~N., et al.\ 2009, \apj, 692, 758 
\bibitem[G{\'o}rski et al.(2005)]{gorski05} G{\'o}rski, K.~M., Hivon, E., Banday, A.~J., et al.\ 2005, \apj, 622, 759 
\bibitem[Greig et al.(2017)]{greig17} Greig, B., Mesinger, A., Haiman, Z., \& Simcoe, R.~A.\ 2017, \mnras, 466, 4239 
\bibitem[Haardt \& Madau(2012)]{haardt12} Haardt, F., \& Madau, P.\ 2012, \apj, 746, 125 
\bibitem[Hewett \& Foltz(2003)]{hewett03} Hewett, P.~C., \& Foltz, C.~B.\ 2003, \aj, 125, 1784 
\bibitem[Hook et al.(2004)]{hook04} Hook, I.~M., J{\o}rgensen, I., Allington-Smith, J.~R., et al.\ 2004, \pasp, 116, 425 
\bibitem[Hopkins et al.(2005)]{hopkins05} Hopkins, P.~F., Hernquist, L., Cox, T.~J., et al.\ 2005, \apj, 630, 705 
\bibitem[Irwin et al.(2004)]{irwin04} Irwin, M.~J., Lewis, J., Hodgkin, S., et al.\ 2004, \procspie, 5493, 411 
\bibitem[Jiang et al.(2007)]{jiang07} Jiang, L., Fan, X., Ivezi{\'c}, {\v Z}., et al.\ 2007, \apj, 656, 680 
\bibitem[Jiang et al.(2016)]{jiang16} Jiang, L., McGreer, I.~D., Fan, X., et al.\ 2016, \apj, 833, 222 
\bibitem[Kashikawa et al.(2015)]{kashikawa15} Kashikawa, N., Ishizaki, Y., Willott, C.~J., et al.\ 2015, \apj, 798, 28 
\bibitem[Kaurov \& Gnedin(2015)]{kaurov15} Kaurov, A.~A., \& Gnedin, N.~Y.\ 2015, \apj, 810, 154 
\bibitem[Kellermann et al.(1989)]{kellermann89} Kellermann, K.~I., Sramek, R., Schmidt, M., Shaffer, D.~B., \& Green, R.\ 1989, \aj, 98, 1195 
\bibitem[Kirkpatrick et al.(2011)]{kirkpatrick11} Kirkpatrick, J.~D., Cushing, M.~C., Gelino, C.~R., et al.\ 2011, \apjs, 197, 19 
\bibitem[Khaire \& Srianand(2015)]{khaire15} Khaire, V., \& Srianand, R.\ 2015, \mnras, 451, L30 
\bibitem[Knigge et al.(2008)]{knigge08} Knigge, C., Scaringi, S., Goad, M.~R., \& Cottis, C.~E.\ 2008, \mnras, 386, 1426 
\bibitem[Koptelova et al.(2017)]{koptelova17} Koptelova, E., Hwang, C.-Y., Yu, P.-C., Chen, W.-P., \& Guo, J.-K.\ 2017, Scientific Reports, 7, 41617 
\bibitem[Kulkarni et al.(2018)]{kulkarni18} Kulkarni, G., Worseck, G., \& Hennawi, J.~F.\ 2018, arXiv:1807.09774 
\bibitem[Illingworth et al.(2013)]{illingworth13} Illingworth, G.~D., Magee, D., Oesch, P.~A., et al.\ 2013, \apjs, 209, 6 
\bibitem[Lang(2014)]{lang14} Lang, D.\ 2014, \aj, 147, 108 
\bibitem[Lawrence et al.(2007)]{lawrence07} Lawrence, A., Warren, S.~J., Almaini, O., et al.\ 2007, \mnras, 379, 1599 
\bibitem[Lusso et al.(2015)]{lusso15} Lusso, E., Worseck, G., Hennawi, J.~F., et al.\ 2015, \mnras, 449, 4204 
\bibitem[Madau et al.(1999)]{madau99} Madau, P., Haardt, F., \& Rees, M.~J.\ 1999, \apj, 514, 648 
\bibitem[Madau \& Haardt(2015)]{madau15} Madau, P., \& Haardt, F.\ 2015, \apjl, 813, L8 
\bibitem[Madau \& Rees(2000)]{madau00} Madau, P., \& Rees, M.~J.\ 2000, \apjl, 542, L69 
\bibitem[Mainzer et al.(2011)]{mainzer11} Mainzer, A., Bauer, J., Grav, T., et al.\ 2011, \apj, 731, 53 
\bibitem[Matsuoka et al.(2018b)]{matsuoka18} Matsuoka, Y., Iwasawa, K., Onoue, M., et al.\ 2018, arXiv:1803.01861 
\bibitem[Matsuoka et al.(2016)]{matsuoka16} Matsuoka, Y., Onoue, M., Kashikawa, N., et al.\ 2016, \apj, 828, 26 
\bibitem[Matsuoka et al.(2018a)]{matsuoka17} Matsuoka, Y., Onoue, M., Kashikawa, N., et al.\ 2018, \pasj, 70, S35 
\bibitem[Mazzucchelli et al.(2017)]{mazzucchelli17} Mazzucchelli, C., Ba{\~n}ados, E., Venemans, B.~P., et al.\ 2017, \apj, 849, 91 
\bibitem[McQuinn et al.(2011)]{mcquinn11} McQuinn, M., Oh, S.~P., \& Faucher-Gigu{\`e}re, C.-A.\ 2011, \apj, 743, 82 
\bibitem[Meisner et al.(2017)]{meisner17} Meisner, A.~M., Lang, D., \& Schlegel, D.~J.\ 2017, \aj, 154, 161 
\bibitem[McGreer et al.(2018)]{mcgreer18} McGreer, I.~D., Fan, X., Jiang, L., \& Cai, Z.\ 2018, \aj, 155, 131 
\bibitem[McGreer et al.(2013)]{mcgreer13} McGreer, I.~D., Jiang, L., Fan, X., et al.\ 2013, \apj, 768, 105 
\bibitem[McGreer et al.(2010)]{mcgreer10} McGreer, I.~D., Hall, P.~B., Fan, X., et al.\ 2010, \aj, 140, 370 
\bibitem[McLeod et al.(2012)]{mcleod12} McLeod, B., Fabricant, D., Nystrom, G., et al.\ 2012, \pasp, 124, 1318 
\bibitem[McMahon et al.(2013)]{mcmahon13} McMahon, R.~G., Banerji, M., Gonzalez, E., et al.\ 2013, The Messenger, 154, 35 
\bibitem[Meiksin (2005)]{meiksin05} Meiksin, A.\ 2005, \mnras, 356, 596 
\bibitem[Miralda-Escud{\'e}(1998)]{me98} Miralda-Escud{\'e}, J.\ 1998, \apj, 501, 15 
\bibitem[Mortlock et al.(2011)]{mortlock11} Mortlock, D.~J., Warren, S.~J., Venemans, B.~P., et al.\ 2011, \nat, 474, 616 
\bibitem[Oke \& Gunn(1982)]{oke82} Oke, J.~B., \& Gunn, J.~E.\ 1982, \pasp, 94, 586 
\bibitem[Page \& Carrera(2000)]{page00} Page, M.~J., \& Carrera, F.~J.\ 2000, \mnras, 311, 433 
\bibitem[Pentericci et al.(2014)]{pentericci14} Pentericci, L., Vanzella, E., Fontana, A., et al.\ 2014, \apj, 793, 113 
\bibitem[Planck Collaboration et al.(2018)]{planck18} Planck Collaboration, Aghanim, N., Akrami, Y., et al.\ 2018, arXiv:1807.06209 
\bibitem[Reed et al.(2017)]{reed17} Reed, S.~L., McMahon, R.~G., Martini, P., et al.\ 2017, \mnras, 468, 4702 
\bibitem[Reichard et al.(2003)]{reichard03} Reichard, T.~A., Richards, G.~T., Hall, P.~B., et al.\ 2003, \aj, 126, 2594 
\bibitem[Richards et al.(2006)]{richards06} Richards, G.~T., Strauss, M.~A., Fan, X., et al.\ 2006, \aj, 131, 2766 
\bibitem[Risaliti \& Elvis(2010)]{risaliti10} Risaliti, G., \& Elvis, M.\ 2010, \aap, 516, A89 
\bibitem[Robertson et al.(2015)]{robertson15} Robertson, B.~E., Ellis, R.~S., Furlanetto, S.~R., \& Dunlop, J.~S.\ 2015, \apjl, 802, L19 
\bibitem[Ross et al.(2012)]{ross12} Ross, N.~P., Myers, A.~D., Sheldon, E.~S., et al.\ 2012, \apjs, 199, 3 
\bibitem[Schindler et al.(2018)]{Schindler18} Schindler, J.-T., Fan, X., McGreer, I.~D., et al.\ 2018, \apj, 863, 144 
\bibitem[Schlegel et al.(1998)]{SFD98} Schlegel, D.~J., Finkbeiner, D.~P., \& Davis, M.\ 1998, \apj, 500, 525 
\bibitem[Schmidt et al.(1989)]{schmidt89} Schmidt, G.~D., Weymann, R.~J., \& Foltz, C.~B.\ 1989, \pasp, 101, 713 
\bibitem[Selsing et al.(2016)]{selsing16} Selsing, J., Fynbo, J.~P.~U., Christensen, L., \& Krogager, J.-K.\ 2016, \aap, 585, A87 
\bibitem[Semelin(2016)]{semelin16} Semelin, B.\ 2016, \mnras, 455, 962 
\bibitem[Shankar et al.(2010)]{shankar10} Shankar, F., Crocce, M., Miralda-Escud{\'e}, J., Fosalba, P., \& Weinberg, D.~H.\ 2010, \apj, 718, 231 
\bibitem[Shen et al.(2007)]{shen07} Shen, Y., Strauss, M.~A., Oguri, M., et al.\ 2007, \aj, 133, 2222 
\bibitem[Simcoe et al.(2008)]{simcoe08} Simcoe, R.~A., Burgasser, A.~J., Bernstein, R.~A., et al.\ 2008, \procspie, 7014, 70140U 
\bibitem[Stevenson et al.(2016)]{stevenson16} Stevenson, K.~B., Bean, J.~L., Seifahrt, A., et al.\ 2016, \apj, 817, 141 
\bibitem[Stocke et al.(1992)]{stocke92} Stocke, J.~T., Morris, S.~L., Weymann, R.~J., \& Foltz, C.~B.\ 1992, \apj, 396, 487 
\bibitem[Tang et al.(2017)]{tang17} Tang, J.-J., Goto, T., Ohyama, Y., et al.\ 2017, \mnras, 466, 4568 
\bibitem[Tsuzuki et al.(2006)]{tsuzuki06} Tsuzuki, Y., Kawara, K., Yoshii, Y., et al.\ 2006, \apj, 650, 57 
\bibitem[Vanden Berk et al.(2001)]{vanden01} Vanden Berk, D.~E., Richards, G.~T., Bauer, A., et al.\ 2001, \aj, 122, 549 
\bibitem[van Leeuwen et al.(2017)]{gaiadr1} van Leeuwen, F., Evans, D.~W., De Angeli, F., et al.\ 2017, \aap, 599, A32 
\bibitem[Venemans et al.(2015)]{venemans15} Venemans, B.~P., Ba{\~n}ados, E., Decarli, R., et al.\ 2015, \apjl, 801, L11 
\bibitem[Venemans et al.(2013)]{venemans13} Venemans, B.~P., Findlay, J.~R., Sutherland, W.~J., et al.\ 2013, \apj, 779, 24 
\bibitem[Venemans et al.(2017)]{venemans17} Venemans, B.~P., Walter, F., Decarli, R., et al.\ 2017, \apjl, 851, L8 
\bibitem[Vestergaard \& Wilkes(2001)]{vestergaard01} Vestergaard, M., \& Wilkes, B.~J.\ 2001, \apjs, 134, 1 
\bibitem[Volonteri \& Rees(2006)]{volonteri06} Volonteri, M., \& Rees, M.~J.\ 2006, \apj, 650, 669 
\bibitem[Volonteri et al.(2015)]{volonteri15} Volonteri, M., Silk, J., \& Dubus, G.\ 2015, \apj, 804, 148 
\bibitem[Wang et al.(2007)]{wang07} Wang, R., Carilli, C.~L., Beelen, A., et al.\ 2007, \aj, 134, 617 
\bibitem[Wang et al.(2017)]{wang17} Wang, F., Fan, X., Yang, J., et al.\ 2017, \apj, 839, 27  (Paper I)
\bibitem[Wang et al.(2015)]{wang15} Wang, F., Wu, X.-B., Fan, X., et al.\ 2015, \apjl, 807, L9 
\bibitem[Wang et al.(2016)]{wang16} Wang, F., Wu, X.-B., Fan, X., et al.\ 2016, \apj, 819, 24 
\bibitem[Wang et al.(2018)]{wang18} Wang, F., et al.\ 2018, \apjl, submitted   (Paper II)
\bibitem[Willott et al.(2010)]{willott10} Willott, C.~J., Delorme, P., Reyl{\'e}, C., et al.\ 2010, \aj, 139, 906 
\bibitem[Wright et al.(2010)]{wright10} Wright, E.~L., Eisenhardt, P.~R.~M., Mainzer, A.~K., et al.\ 2010, \aj, 140, 1868 
\bibitem[Wu et al.(2015)]{wu15} Wu, X.-B., Wang, F., Fan, X., et al.\ 2015, \nat, 518, 512 
\bibitem[Wyithe \& Loeb(2003)]{wyithe03} Wyithe, J.~S.~B., \& Loeb, A.\ 2003, \apj, 595, 614 
\bibitem[Yang et al.(2016)]{yang16} Yang, J., Wang, F., Wu, X.-B., et al.\ 2016, \apj, 829, 33 
\bibitem[York et al.(2000)]{york00} York, D.~G., Adelman, J., Anderson, J.~E., Jr., et al.\ 2000, \aj, 120, 1579 
\bibitem[Yuan et al.(2013)]{yuan13} Yuan, H., Zhang, H., Zhang, Y., et al.\ 2013, Astronomy and Computing, 3, 65 
\bibitem[Zou et al.(2017)]{zou17} Zou, H., Zhang, T., Zhou, Z., et al.\ 2017, arXiv:1712.09165 


\end{thebibliography}
\end{document}